\documentclass[sigconf]{acmart}
\makeatletter                   
\def\mdseries@tt{m}             
\makeatother                    
\usepackage[plain]{fancyref}
\usepackage[draft=true]{minted} 
\usepackage{color}
\usepackage{hyperref}           
\hypersetup{
    colorlinks=true,
    linkcolor=blue,
    filecolor=red,      
    urlcolor=magenta,
    breaklinks=true,            
}
\usepackage{breakurl}           

\AtBeginDocument{%
  \providecommand\BibTeX{{%
    \normalfont B\kern-0.5em{\scshape i\kern-0.25em b}\kern-0.8em\TeX}}}

\definecolor{shmgreen}{RGB}{126, 166, 61}

\newcommand {\bjoern}[1]{ }
\newcommand {\jd}[1]{ }
\newcommand {\fred}[1]{ }
\newcommand {\shm}[1]{ }

 \newcommand{\edits}[1]{#1}
 \newcommand{\deletes}[1]{}

\definecolor{OliveGreen}{HTML}{3C8031}
\definecolor{Fuchsia}{HTML}{8C368C}

\usepackage{verbatim}
\usepackage{xspace}
\newcommand{\tool}[0]{\textsc{DreamSheets}\xspace}

\copyrightyear{2024}
\acmYear{2024}
\setcopyright{rightsretained}
\acmConference[CHI '24]{Proceedings of the CHI Conference on Human Factors in Computing Systems}{May 11--16, 2024}{Honolulu, HI, USA}
\acmBooktitle{Proceedings of the CHI Conference on Human Factors in Computing Systems (CHI '24), May 11--16, 2024, Honolulu, HI, USA}
\acmDOI{10.1145/3613904.3642858}
\acmISBN{979-8-4007-0330-0/24/05}

\begin{document}
\sloppy                         
\title[DreamSheets]{Prompting for Discovery: Flexible Sense-Making for\\AI Art-Making with DreamSheets}




\author{Shm Garanganao Almeda, J.D. Zamfirescu-Pereira, Kyu Won Kim, \\
Pradeep Mani Rathnam, Bjoern Hartmann}
\affiliation{%
  \institution{UC Berkeley}
  \city{Berkeley}
  \state{CA}
  \country{USA}
}



\renewcommand{\shortauthors}{Almeda, Zamfirescu-Pereira, Kim, Mani Rathnam, and Hartmann}

\begin{abstract}

Design space exploration (DSE) for Text-to-Image (TTI) models entails navigating a vast, opaque space of possible image outputs, through a commensurately vast input space of hyperparameters and prompt text. Perceptually small movements in prompt-space can surface unexpectedly disparate images. How can interfaces support end-users in reliably steering prompt-space explorations towards interesting results?
Our design probe, DreamSheets, supports user-composed exploration strategies with LLM-assisted prompt construction and large-scale simultaneous display of generated results, hosted in a spreadsheet interface. 
Two studies, a preliminary lab study and an extended two-week study where five expert artists developed custom TTI sheet-systems, reveal various strategies for targeted TTI design space exploration---such as using templated text generation to define and layer semantic ``axes'' for exploration. We identified patterns in exploratory structures across our participants' sheet-systems: configurable exploration ``units'' that we distill into a UI mockup, and generalizable UI components to guide future interfaces.

\end{abstract}




\begin{teaserfigure}
  \centering
  \includegraphics[width=.9\linewidth]{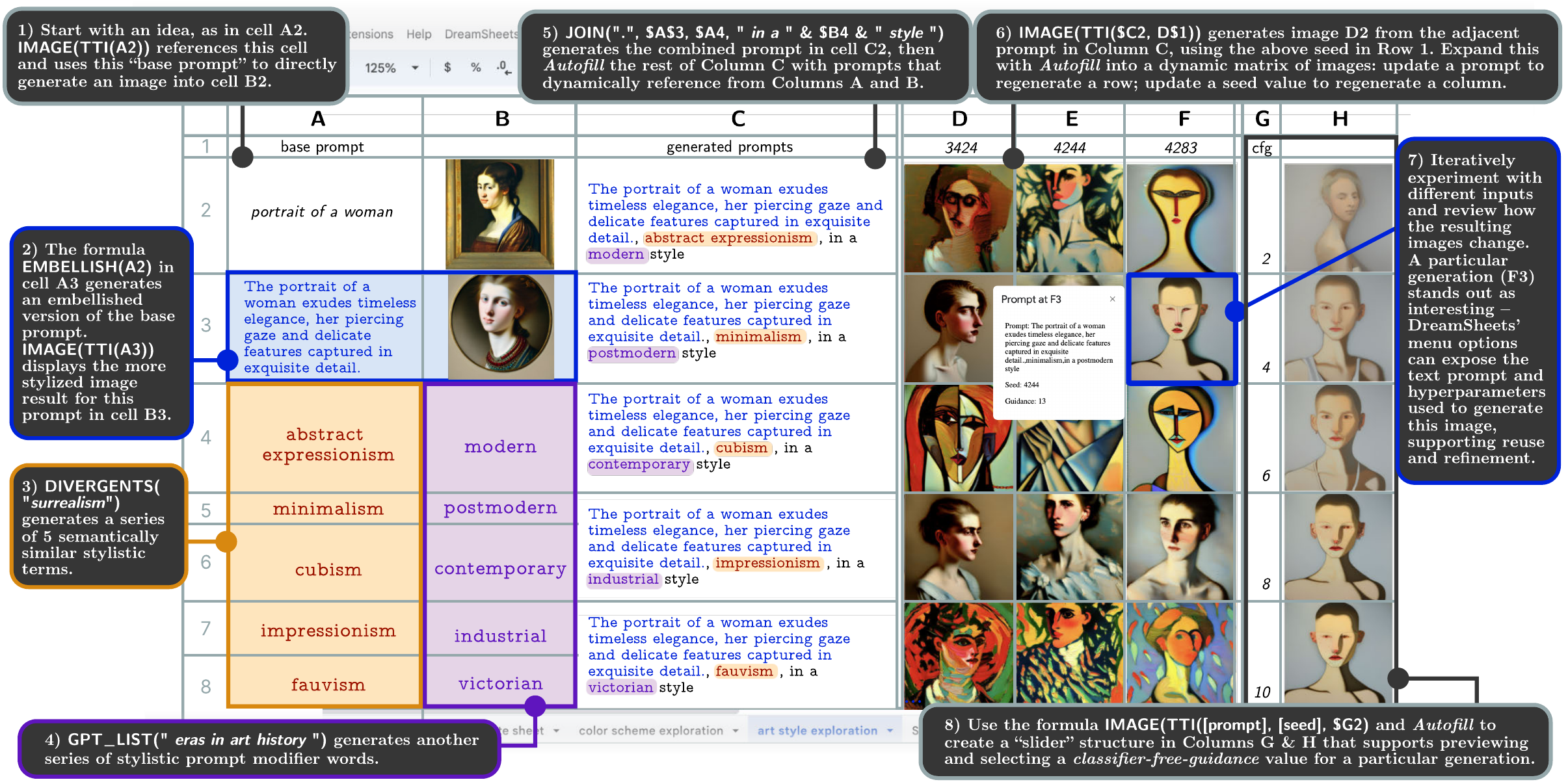}
  \caption{ \tool provides a 2D spreadsheet interface that Text-to-Image users can use to author creative workflows for rapid prompt-image exploration. In this example derived from sheet-systems authored by expert participants (E1, E2, E4, E5) an initial idea for a prompt (A2) is used to generate an image (B2) using \tool's \textsc{TTI()} function. They modify the prompt with \textsc{EMBELLISH()}, an LLM-based function, to generate the more stylized (B3). Other LLM functions \textsc{DIVERGENTS()} and \textsc{GPT\_LIST()} are used to generate series of \textit{art styles} and \textit{eras in art history}, two chosen axes for prompt exploration. Their dynamic \textit{prompt template} combines the embellished prompt (A3) with references to the generated lists of ``modifiers'' to fill Column C with a series of prompts. These dynamic prompts generate the images in Columns D-F. Each image column utilizes a different seed, referencing the cell in row 1. Columns G and H depict a hyperparameter ``slider'' structure used to evaluate and select a cfg setting. If content in a referenced cell is updated, the sheet automatically regenerates dependent cells, allowing users to repurpose this workflow structure for iterative explorations.
} 
  \label{fig:teaser}
  \Description{This teaser figure features Dreamsheets’ 2D spreadsheet interface that is housed within Google Sheets; rows are numbered, and each column has a letter from A-H. The top-left-most cell, A1, contains the text “base prompt” and the corresponding text, “portrait of a woman” is in cell A2 below. This prompt is used to generate an image of a woman in a picture frame in cell B2. Below, a more detailed version of the prompt in cell A3, generated using the EMBELLISH() LLM function: “The portrait of a woman exudes timeless elegance, her piercing gaze and delicate features captured in exquisite detail.” Adjacent to the more detailed prompt is the corresponding image generation in cell B3, a more elegant and apparently detailed portrait of a woman. In the rest of column A, in cells A4-A8, there is a series of words: “abstract expressionism”, “minimalism”, “cubism”, “impressionism”,”fauvism”. Another list of words in the adjacent cells B4-B8:"modern", "postmodern", "contemporary", "industrial", "victorian". these lists are generated with LLM functions DIVERGENTS(“surrealism”) and GPT_LIST(“eras in art history”) respectively. The has authored a dynamic prompt template that concatenates the embellished base prompt (cell A3) with one word from each of the 2 lists of “modifiers” generated by the LLM functions, filling Column C with a series of prompts: “The portrait of a woman exudes timeless elegance, her piercing gaze and delicate features captured in exquisite detail, abstract expressionism, in a modern style,” in C2, “The portrait of a woman exudes timeless elegance, her piercing gaze and delicate features captured in exquisite detail, minimalism, in a postmodern style,” in C3, and so on across column C.  They have used these dynamic prompts to generate a matrix (3 cells wide, 5 cells high) of images in Columns D-F. Each row corresponds to one of the dynamically combined prompts. The first row contains portraits of a woman in an abstract expressionist+modern style; they look painterly and somewhat abstract compared to the original generated image. Each row has a different painting style corresponding to the respective “modifier” words that contributed to its prompt. Each column has a number in the top cell (in row 1): 3424, 4244, and 4238. These indicate the different seed for each row; since the seed changes the initial random noise used to generate the images, each column has some similarities in terms of composition and color; the first row primarily features women with a side profile facing the right, for instance. In this way, there are 3 variations to each prompt, for each of 5 prompts. There is a blue box around image F3, a portrait of a feminine character with a buzzed haircut, in a painterly, simple “minimalist and postmodern” style, indicating that the user has selected this generation as interesting. The prompt used to generate F3 is exposed using DreamSheets menu options. Column H shows a series of variations on this one generation; in H2, a blurry portrait of the bald character, which progressively gets more detailed and higher contrast with minor other image differences in cells H3-H8. Columns G and H show a “context-free-guidance” slider structure; G is a series of values, which are used to modify the F3 generation and produce the series of progressively different images in column H.}
\end{teaserfigure}
\maketitle

\section{Introduction}

Text-to-Image (TTI) models like DALL•E~\cite{ramesh22} and Stable Diffusion~\cite{rombach22} generate images from a combination of text prompts and numerical parameters (e.g., seeds). Interfaces for these models are seeing wide end-user adoption in a variety of settings, from marketing collateral to independent art making. 

A critical skill for Text-to-Image (TTI) users is navigating between prompt text inputs and image outputs. Building such an understanding is neither trivial nor straightforward~\cite{zamfirescu23why,zamfirescu23herding}: the spaces of possible inputs and outputs are \emph{massive}, and the mapping of one to the other is highly \emph{opaque}.

Current patterns in commercial interfaces present limited support for exploration: a text box for prompt input and an area for displaying and saving a few output images. Some offer a number of additional features to support \textit{prompt engineering}: canned prompt ideas, including options to influence ``style,'' and sliders for manipulating hyperparameters. 

This lack of explicit interface support has led to the creation of resources such as community-curated prompt books~\cite{parsons2022thedalle}, spreadsheets~\cite{promptparrot}, and tutorials~\cite{Nielsen2022,salerno22how} 
that document exploration processes. Researchers have also begun to study prompting practices~\edits{\cite{kulkarni_word_2023,ChilltonCHI22-prompt-guideline-txt-img, chang_prompt_2023,oppenlaender2022creativity,oppenlaender2023cultivated,oppenlaender2023taxonomy}} and to propose alternative interfaces for interacting with TTI (e.g., Promptify~\cite{brade2023promptify}) and other types of generative models (e.g., GanZilla~\cite{evirgen2022ganzilla}, Spacesheets~\cite{loh2018spacesheets}, \edits{PromptAid~\cite{mishra2023promptaid}}). These systems tend to support particular prompt-image or prompt-text workflows, helping users refine prompts towards a goal. 

In this paper we argue that supporting Text-To-Image users goes beyond ensuring that they can achieve a particular end result; \textit{gaining an understanding of the mapping between input and output is core to successful creation with generative AI systems}. 
Exploring the relationship between TTI inputs and outputs is a sensemaking process~\cite{pirolli2005sensemaking,russell1993sensemaking} where users aim to build mental representations that allow them to reliably navigate to desired outputs. While the input and output spaces are massive and opaque, they are \textit{not} arbitrary: prompters \textit{can} develop reliable exploration ``targeting strategies'' with experience and productive reflection. Users can use the knowledge they gain from many iterative input-output tests to effectively  ``prompt-craft,'' effectively steering generations towards desirable results.

Thus, our guiding research question is: \emph{How might new interfaces support users in sense-making for successful art making with such models?}

To investigate this question, we built \tool, a tool that enables TTI model users to compose \textit{targeted exploration} systems within a spreadsheet interface. In \tool, spreadsheet cells can contain prompts, or images generated from those prompts. A set of novel prompt manipulation functions enable users to explore prompt space computationally, through the construction and strategic combination of categorical lists, alternative wordings, embellishments, synonyms, and more. These functions are implemented through prompts to a large language model (LLM).

Spreadsheets may not readily provide the ideal affordances for organizing image collections; however, they are a \textit{highly flexible, computational} substrate for \textit{what-if} exploration; by presenting image and text generation tools within a customizable sandbox, \tool enables users to compose novel creative workflows.

We investigated users' exploration strategies in two studies with \tool: a 1-hour lab study with 12 primarily amateur participants, and a two-week extended study with five expert TTI artists. In these studies, we examined how both groups (1) develop intuition for prompt designs that yield specific outputs, and (2) use \tool' affordances for computational prompt manipulation, workflow creation, and output evaluation.

Our primary insights lie in observing and analyzing how participants used \tool to develop custom TTI sheet-systems, and in identifying sense-making and exploration patterns across these generative prototypes, including the construction and iterative reuse of composable exploration ``structures'' that map to generalizable UI concepts.
We then use these insights to create UI mockups to inform potential future interfaces, and report on the feedback and speculation it elicited from our participants.

This paper makes three contributions: 

First, it describes \tool, a spreadsheet-based TTI design space exploration platform that enables user-defined computational and LLM-supported interactions over the joint design space of prompts, seeds, and other TTI model hyperparameters.

Second, it offers a rich description of a first-of-its-kind extended (2-week) study exploring the ways in which artists use computational structures for sense-making and to explore the design space of TTI models. 

Finally, it presents a set of generalized UI design suggestions, co-designed in a visual UI mock-up with our expert artist participants, and informed by the sheet-systems they developed while working in \tool.

\section{Related Work}

Our work builds on research showing that considering many alternatives in parallel effectively aids design space exploration ~\cite{hartmann-juxtapose-2008,rawn2023understanding}, such as through gallery interfaces~\cite{marks1997design}, tracking exploration history~\cite{lee2010designing,heer2008graphical}, with suggestions for possible input shifts~\cite{marks1997design}, and effective organization~\cite{woodbury_burrow_typology_2006,ritchie2011dtour,lunzer-subjunctive-tochi08}. Spreadsheets enable many of these abilities that common TTI interfaces often lack.

In this section, we draw explicit connections to Creativity Support Tools, prompting and other TTI model workflows, design space exploration of images in non-TTI contexts, and sensemaking.





\subsection{Creativity Support}

In 2007, Shneiderman identified four underlying design principles for creativity support tools (CSTs): \textit{support exploratory search}, \textit{enable collaboration}, \textit{provide rich history-keeping}, and \textit{design with low thresholds, high ceilings, and wide walls}~\cite{shneiderman02}. A more recent body of research explores how CST design can aid users' creative processes and productivity \cite{Frich19, young21}. Spreadsheets are themselves an example of a tool that supports creative exploration, enabling users to separate fixed values from values they want to vary, affording effective exploration and evaluation of ``what-if'' scenarios ~\cite{shneiderman02, resnick05}.

\subsection{Prompting \& Text-to-Image Model Workflows}


At the surface, prompting can appear straightforward, but crafting effective prompts is a challenge, \edits{even for experts}~\cite{zamfirescu23why,zamfirescu23herding,Liu22}. How a prompt directly impacts model outputs is an active area of research~\cite{liu2021pretrain,T0-Sanh2021}. Choosing the right language to achieve desirable visual results in these prompt-based interactions can be difficult, motivating online user communities~\cite{parrotzone,Durant2023} 
and researchers to develop and investigate new prompting techniques~\cite{Liu22, liu2023dalle} and tools supporting prompt discovery and exploration~\cite{serra22, brade2023promptify}. These tools tend to be goal-driven, helping artists with a particular image ``goal'' target and improve their generations, while perhaps considering alternatives. In contrast, \tool seeks to explicitly support the rapid construction of flexible structures towards various user-defined goals. \edits{A growing body of literature has also begun to explore how communities of practice are approaching prompting~\cite{oppenlaender2023cultivated, oppenlaender2023taxonomy, chang_prompt_2023, oppenlaender2023prompting}. These studies offer taxonomies of prompt structures, showing how artists use different prompt modifiers~\cite{oppenlaender2023taxonomy}, how they engage in the process of prompt engineering~\cite{oppenlaender2023prompting}, and how they consider the material properties of text prompts~\cite{chang_prompt_2023}; this literature will rightfully continue to grow as these practices evolve alongside the models they rely on.}

\edits{\tool builds upon this prior work by enabling prompt-artists to rapidly prototype computational structures and pursue user-defined, targeted explorations across prompt-image space---allowing us, in turn, to learn how artists use such new capabilities to build structures for sense-making and art-making.}



% ...



\subsection{Design Space Exploration of Images}
\label{sec:design-space-exploration}
Prompt discovery tools follow a long line of research into design space exploration of images. 
In spaces like computer graphics and animation where \textit{visual judgment} of a human-designer-produced artifact is the primary evaluation mode, prior work has often focused on browsing interfaces, such as in Marks \textit{et al.}'s seminal Design Galleries~\cite{marks1997design}. 
Interaction techniques for browsing include multi-step galleries~\cite{marks1997design}; map metaphors~\cite{talton-exploratory-2009}; or faceted browsing~\cite{glassman-examplore-2018}. 

Narrowing down from the explored designs, users may wish to pursue multiple alternative options for deeper exploration, though typically many orders of magnitude fewer than the number of algorithmically explored designs, as in GEM-NI~\cite{zaman-gemni-chi15}. 

Spreadsheets' usefulness for visual design space exploration in part stems from the intrinsic 2D matrix layout enabling ``small multiples'', a term Edward Tufte popularized~\cite{tufte1993envisioning} as an answer to the question ``compared to what?'' In a ``contact sheet''-like 2D matrix, one can readily compare many images at once and quickly identify the best candidates. Spreadsheets have a rich history of serving as vehicles for exploratory work, in accounting and far beyond--–utilized as early as 1994 for information visualization of data and images themselves~\cite{chi97spreadsheet,levoy1994spreadsheets}, including images generated from a numerical input space~\cite{loh2018spacesheets}.

\subsection{Sense-making}

A number of our observations relate to the broader sense-making literature, including Pirolli and Card's seminal work on information foraging~\cite{pirolli1999information}---\tool offers users an \textit{information scent} on prompts---and sense-making more broadly~\cite{pirolli2005sensemaking,russell1993sensemaking}. This line of work models how users navigate and make decisions in information-rich environments (like \tool), balancing between the perceived cost of seeking information and the potential reward of finding what they're seeking. \tool's design draws upon the \textit{free energy principle of the brain} theory from cognitive science~\cite{friston2009free} which describes how the brain reduces uncertainty by making predictions and updating an internal mental model accordingly, generatively optimizing its internal model with sensory input to enhance prediction accuracy. This principle formed a basis for Davis et.al's Creative Sense-Making (CSM) framework~\cite{davis_csm}, which they applied to human-AI co-creation in the collaborative drawing domain. \tool's design also draws inspiration and lessons from existing sense-making interfaces – including classics like Scatter/Gather~\cite{cutting2017scatter} and more modern implementations like Sensecape~\cite{suh2023sensecape}. \edits{SemanticCollage~\cite{koch2020semantic} and ImageSense~\cite{koch2020imagesense} provided reusable, system-generated text and visuals to support creative image search and sense-making with ``reflection in action''.}

\section{Prompt \& Image Exploration with \tool}
\tool leverages the inherently flexible spreadsheet model in support of iteration and exploration of the TTI generation prompt input space.  
The features of \tool are embedded within a spreadsheet (built on Google Sheets) that recomputes and re-renders images \edits{in response to prompt additions and changes, allowing (for instance) drag-based ``autofill'' of columns, rows, or 2D regions with formula-generated prompts and images based on those prompts, alongside other common spreadsheet functionality. Specifically, }\tool offers access to diffusion model image generation as a spreadsheet function that can take the content of other cells in the sheet as input, including combinations or transformations of multiple cells. These features support the user in efficiently exploring, observing how generated outputs are influenced by modifications to the input. 

Our prototype also includes a set of LLM-based spreadsheet functions for manipulating prompts directly, such as \textsc{gpt\_list} and \textsc{list\_completion} for generating or extending a list of items of a certain description, \textsc{embellish} to create a detailed variation of  the input text, and \textsc{alternatives} to generate multiple variations of a seed prompt (see Table~\ref{tab:functions} for a full list). \edits{These functions serve as an aid to users in the construction of axes along which to explore the prompt/image design space, supporting template prompts with automatic value insertion. 

For example, a template prompt like ``A \textit{<facial expression>} woman'' can be expanded into a column of different prompts by substituting values generated by \textsc{gpt\_list("facial expressions")}.}

\begin{table}[t] 
\centering
\renewcommand*{\arraystretch}{1.2} 
\begin{tabular}{p{0.38\linewidth} p{0.53\linewidth}} 

\toprule[2pt]
\textbf{Function Name} & \textbf{Description}\\
\midrule[2pt]
\textsc{tti}(\textit{prompt, [seed], [cfg]}) & Generate an image (at the returned URL) using the given \textit{prompt} and, optionally, \textit{seed} and a classifier-free guidance (\textit{cfg)} parameters. \\
\hline
\textsc{gpt}(\textit{prompt}) & LLM function for arbitrary \textit{prompt}.\\
\hline
\textsc{gpt\_list}(\textit{prompt, length}) & Populates \textit{length} cells in a row (or column) with words/phrases of type \textit{prompt}.\\
\hline
\textsc{list\_completion}(\textit{prompt}) & Like \textsc{gpt\_list}, but \textit{prompt} is a list of items, rather than a description.\\
\hline
\textsc{synonyms}(\textit{prompt}) & Generates a list of synonyms.\\
\hline
\textsc{antonyms}(\textit{prompt}) & Generates a list of antonyms.\\
\hline
\textsc{divergents}(\textit{prompt}) & Generates ``divergent'' words.\\
\hline
\textsc{alternatives}(\textit{prompt}) & Generates a list of alternative wordings for \textit{prompt}.\\
\hline
\textsc{embellish}(\textit{prompt}) & Generates an embellished alternative to \textit{prompt}, commonly using more specific or detailed words.\\
\bottomrule[2pt]
\end{tabular}
\vspace{0.1cm}
\caption{The LLM-based functions available in our prototype. Each list-producing function additionally has an \textsc{\_t} alternative (e.g., \textsc{synonyms\_t}) that transposes its outputs across a row.}
\label{tab:functions}
\Description{A list of all of the new functions added by \tool, and their functionality.}
\end{table}

\subsection{\tool Design}

\edits{Our primary goal here was to enable the use of a computational substrate for TTI model design space exploration via spreadsheet formula construction.}

\deletes{From a design perspective, integrating TTI model functionality was relatively straightforward---we devoted significant design energy to developing a set of prompt exploration functions (see Table~\ref{tab:functions}), themselves based on prompting a LLM.}

Drawing on prior work in prompt design~\cite{oppenlaender2022creativity, oppenlaender2023taxonomy, ChilltonCHI22-prompt-guideline-txt-img,parsons2022thedalle}, we identified the testing of alternative phrasings and the addition of detail as core activities in TTI prompt exploration. These activities help users explore neighboring points in design space and recognize fruitful \textit{directions for further} prompt explorations. \edits{We learned from prior evaluations of prompt engineering in the TTI context~\cite{oppenlaender2023prompting} that users were likely to express a diversity of design patterns and that supporting flexibility, providing immediate visual feedback, and offering an extended period of familiarization would be critical.}

We operationalized support for these exploratory activities as the \textsc{alternatives}, \textsc{divergents}, and \textsc{embellish} functions. Similarly, synonym and antonym generation are core NLP building blocks, useful for creating variation that targets specific words in a longer prompt---we integrated these capabilities through the \textsc{synonyms} and \textsc{antonyms} functions; \edits{see Table~\ref{tab:functions} for a full list. These operations formed a foundation from which users could build their own custom workflows and strategic approaches to discovering model capabilities and exploring the underlying prompt design space.}

To use these concepts in a spreadsheet paradigm and support the generation of \textit{sets} of images, we designed these functions to output \textit{lists} of values that populate across a column (or row) of cells. These terms in these cells can be referenced in traditional spreadsheet style and concatenated with other values to form combined prompts.  We also provided functions to \textit{extend} lists of prompts or prompt parts, allowing users to build on a conceptual list by providing a few initial examples. 

\subsection{\tool Implementation}
\begin{figure*}
    \centering
    \includegraphics[width=0.9\textwidth]{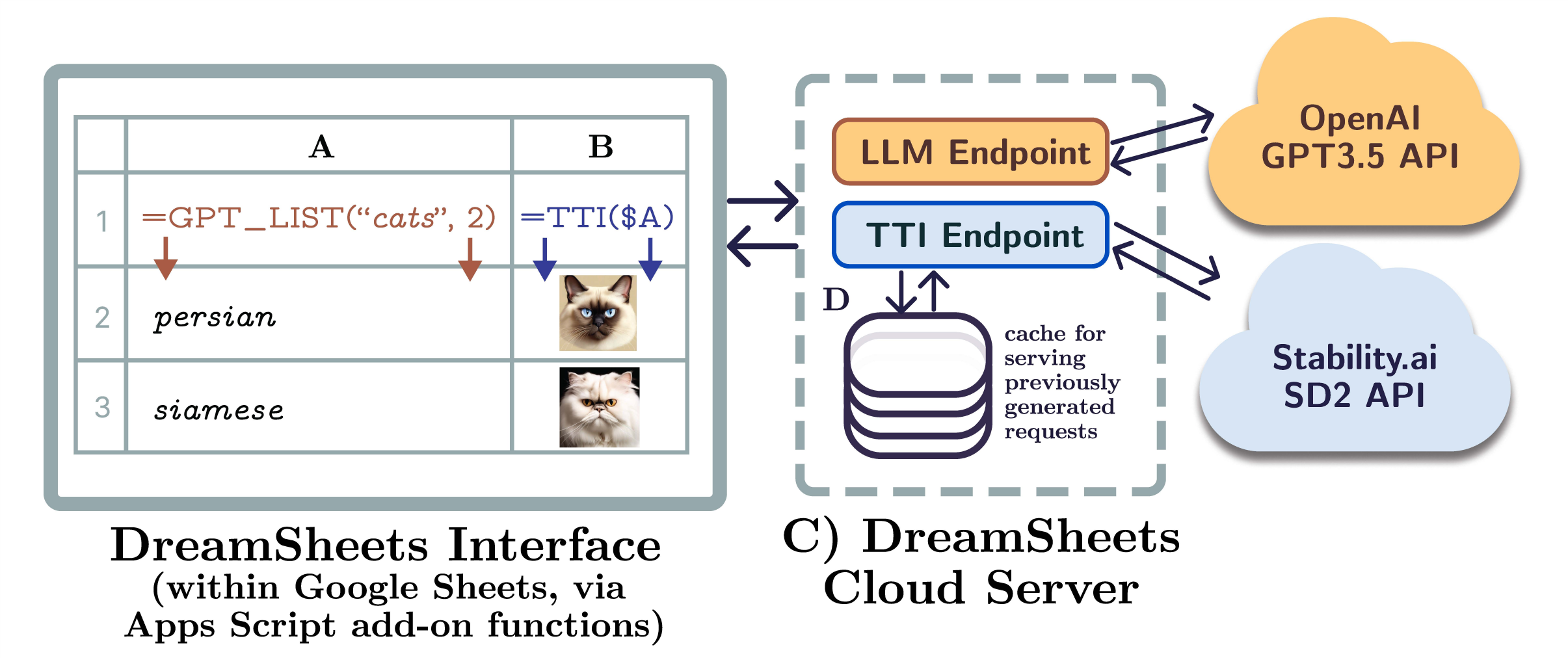}
    \caption{The \tool implementation. LLM (A) and TTI (B) functions fetch from separate endpoints of the \tool cloud server (C) which forwards requests to the OpenAI ChatGPT and Stability.ai Stable Diffusion cloud-based APIs. TTI requests are cached (D) using a hash of (prompt text, seed, classifier-free guidance) as a key.}
    \label{fig:implementation}
    \Description{ A system diagram showing 3 main parts - the Dreamsheets user interface, the DreamSheets cloud server, and 2 clouds labeled OpenAI GPT 3.5 API and Stability.ai SD2 API. DreamSheets User Interface shows that the user has utilized the LLM function “GPT_LIST(“cats”,2)” in column A to generate 2 “cat” words below: persian, siamese. The  TTI function has been used in column B (TTI(\$A)) to generate an image of a persian cat and a siamese cat. Arrows show how these requests are sent to the DreamSheets Cloud Server, depicted as a box consisting of the LLM Endpoint, the TTI Endpoint, and a cache for serving previously generated requests.Arrows show how the LLM endpoint requests and receives from the OpenAI GPT 3.5 API cloud service, and the TTI Endpoint requests and receives from the Stability.ai SD2 API cloud service. The DreamSheets Cloud server is then able to send generated text and images back to the DreamSheets interface within Google Sheets, allowing users to generate text and images and display them directly in spreadsheet cells. }
\end{figure*}

We explored various service options for \tool's underlying spreadsheet functionality, including building our own spreadsheet interface from scratch, open source spreadsheets HandsOnTable\footnote{https://handsontable.com} and LuckySheet,\footnote{https://github.com/dream-num/Luckysheet} and both Microsoft Excel and Google Sheets.

One major challenge in integrating with an existing spreadsheet is the relatively long latency of image generation itself: up to 15 seconds or more, even when using cloud APIs. Spreadsheet users are accustomed to rapid updates and recomputations in response to changes in cell values---a multi-minute delay resulting from a backlog of image prompt updates and, consequently, new image generations, would be unacceptably slow. This need drove our use of the Stability.ai API,\footnote{https://api.stability.ai/docs} which supports parallel image generation requests with the Stable Diffusion 2 model, and offers sub-15 second response times. This critically enabled the full-scale ``small multiples'' visualizations of results that we wanted users to be able to utilize to view and evaluate results across multiple input axes simultaneously.

Ultimately, we selected Google Sheets as the spreadsheet interface, as it is easily extensible and accessible to most people. Google Sheets' Apps Script environment lets developers create add-ons in a JavaScript-like environment, has a sufficiently long timeout (30 seconds) for custom functions, and allows users to continue to edit the sheet even while our custom formulas, which required back-end calls to TTIs and LLMs, awaited responses. 

As a side benefit, because Google Sheets is already an online-native platform, rapid collaboration and version history are built-in. 

We implemented \tool as a Google Sheets Apps Script add-on and a proxy web server written in JavaScript using ExpressJS. The add-on adds custom functions described in Table~\ref{tab:functions}, making the corresponding requests to the proxy web server which handles caching and calling the appropriate API to either Stability.ai or OpenAI. Figure~\ref{fig:implementation}
illustrates how the proxy server facilitates communication between the Google Sheets add-on and Stability.ai or OpenAI. For the \textsc{tti} function, the proxy server makes a hash using a combination of the prompt, seed, and guidance values and checks if the image has been generated before. Otherwise, an API call is made to Stability.ai to generate a $512 \times 512$-pixel image which is then cached in the file system for easy retrieval in the future.

The LLM-based functions that return a list utilizes OpenAI's ChatGPT with \texttt{gpt-3.5-turbo}. To ensure that ChatGPT returns a properly formatted list with the appropriate length, it is initialized with the following messages:
\\

\begin{verbatim}
system: Respond with a Javascript array literal with the
        given length in parentheses
user: types of animals (length: 5)"
assistant: ["dog", "cat", "frog", "horse", "deer"]
user: [PROMPT] (length: [LENGTH])
\end{verbatim}
\vspace{0.75em}
The implementation for each LLM-based function differs only in the prompt sent to the LLM proxy server: each function prepends different additional instructions to the user's inputted prompt. The complete list of full prompts sent to ChatGPT are:

\begin{description}
    \item[\textsc{list\_completion}] \texttt{Similar items to this list without repeating "[LIST]"}
    \item[\textsc{synonyms}] \texttt{Synonyms of "[USER INPUT]"}
    \item[\textsc{antonyms}] \texttt{Antonyms of "[USER INPUT]"}
    \item[\textsc{divergents}] \texttt{Divergent words to "[USER INPUT]"}
    \item[\textsc{alternatives}] \texttt{Alternative ways to say "[USER INPUT]"}
    \item[\textsc{embellish}] \texttt{Embellish this sentence: [USER INPUT]}
\end{description}

\edits{
\subsection{Prototyping Interactions with Spreadsheets}
\label{sec:prototypinginteractions}


In addition to serving as a platform for TTI exploration and sense-making, \tool is a demonstration of a kind of creativity support tool for composable prototyping that can be built on top of existing spreadsheet software as a computational substrate. Spreadsheets are an excellent foundation for tools that might benefit from 2D structure, formula construction, and a familiarized user base---and they can be straightforward to extend if there is alignment between the spreadsheet and backend data models. Here we describe how lessons from \tool can be used in the construction of other TTI-based CSTs.

First, in designing a system with a spreadsheet substrate, a number of considerations arise, grounded in \textit{how much} and \textit{what kind} of customization is needed. Is it sufficient, for example, to simply add new spreadsheet formula functions---or are more substantial changes needed to the interaction layers? Will custom data types need to be supported, and how will they be rendered into cells? How will users interact with these different data types---are new input mechanisms (i.e., beyond text-in-cell input) needed too, such as file uploads or image editing? What response latency is acceptable given the tasks users are expected to engage in? And, finally, how critical are history-keeping and real-time multiplayer functionality?

These considerations constrain which underlying spreadsheets can be used. For \tool, 
we found that Google Sheets' API support was sufficient: custom functions were easy to add, and meeting the required maximum latency (30s) was possible by parallelizing requests to our backend. One major drawback was that custom functions could not cause cells to render images---only the built-in \textsc{=image(\textit{image-url})} formula can do that---forcing the rather clunky \textsc{=image(tti(\textit{prompt, seed, cfg}))} construction. 

In contexts where interaction layer changes are needed, open-source spreadsheets such as HandsOnTable or LuckySheet offer more expansive opportunities for customization. Both options support extending spreadsheets with additional content types for cells as well as supporting additional functionality for user input and manipulation of images, poses, embeddings, and other data types relevant to TTI models. For \tool, this flexibility would have allowed us to build a \textsc{=tti} function that displays an image directly, without needing an intermediate \textsc{=image} function call---as well as integrate image and other types of inputs.

Custom data types can pose challenges too: for opaque (non-user-interpretable) data types, like embeddings, researchers should consider how to represent those values in a sheet itself. In some cases, simply storing the data on a server and sending a unique ID (mapped to each value on said server) would be preferable to passing blob data around a spreadsheet's cells. Similarly, for compute-intensive operations like image synthesis, caching is critical: spreadsheets re-render frequently and unpredictably, and generating an entire sheet's worth of images can be both expensive and time-consuming. 

In \tool, we cached images on the server and used a unique ID to pass around in the browser that represents a specific (prompt, seed, cfg) tuple. This approach could be generalized to support many other types of inputs and outputs, including vectors, embeddings, tokens, and other kinds of state, as well as intermediating models (LoRAs, ControlNet, etc.). Storing data on the server and referencing it by ID allows for less client-side overhead load in a spreadsheet (and potentially cost benefits through caching across users), and makes caching of those values straightforward, but requires an additional translation layer between sheet and backend. 

}

\section{Method}
Having built \tool, we first ran a preliminary 1-hour lab study with novice users to understand how users approach using \tool for TTI exploration; this revealed that \tool enables a variety of custom workflows, but that sensemaking was nearly always among the first activity participants engaged in.

Following our lab study, we ran a second, 2-week extended study with expert users. This second study was intended explicitly to better understand the kinds of custom workflows experts would build for sensemaking, given enough time, and what kinds of individual activities those sensemaking workflows consisted of.

\subsection{Preliminary Lab Study}

In this initial study, we observed how participants used \tool to define and use a text-to-image generation workflow; we gave participants a concrete task and training in the tool, but did not direct them beyond that. 
\subsubsection{Participants}We recruited 12 participants via email lists and social media. 
%
All 12 reported some spreadsheet experience, with most (10 out of 12) reporting frequent use (many times or daily). 10 out of 12 participants also had some experience with TTI models, and only 1 participant (P1) reported no prior experience with LLMs.


\subsubsection{Task and Protocol }
Each study took place through a Zoom call that lasted approximately 60 minutes, during which participants and the facilitating researcher collaborated in a shared Google Sheets document. 
We designed a \textit{concept art} creation task to give users a direction achievable in the short amount of time provided, while leaving room for subjectivity and creativity: users were given a single inspirational image, then asked to generate three new images that could fit in a style and unspecified narrative as suggested by the inspiration image. 
We included a prompt explaining the task in the activity sheet, as well as an image of a post-apocalyptic, ruined Seattle, complete with space needle.\footnote{The image used for this task was borrowed with gratitude from Andy Salerno, whose blog post~\cite{salerno22how} originally inspired this work.} \shm{I commented out the DreamSheets in use figure; should we still include this space needle picture? (see Fig.~\ref{fig:usage}, cell B3).} 


Our protocol began with a brief tutorial on \tool and its functionalities, followed by an observation of participants as they engaged in the concept art task.
We used an example sheet 
to walk participants through a tutorial to first remind participants of general spreadsheet operations (i.e. using formulas with cell references and expanding them with autofill) and then  introducing \tool's image and text generation functions.
%
%
%
Once users were comfortable using the TTI and GPT functions, we introduced the concept art activity. Participants were encouraged to think aloud as they generated the three images required to complete the task.  

\subsubsection{Data Collection and Analysis}
We observed, recorded, and transcribed video and audio of each interview, including approximately 40 minutes of system use, in entirety, throughout which participants were encouraged to think aloud and provide clarification when prompted.
We then engaged in an exploratory qualitative data analysis using the recordings, transcripts, and resulting spreadsheet artifacts.
We recorded responses to surveys completed before and after the interview, and reviewed usage data, containing logs of each text or image generation function call used in \tool. 
\subsubsection{Preliminary Study Results: \tool in use}
Here, we provide an overview of some results specific to the preliminary lab study that informed our second, extended study.
\begin{figure}
    \centering
    \includegraphics[width=0.90\linewidth]{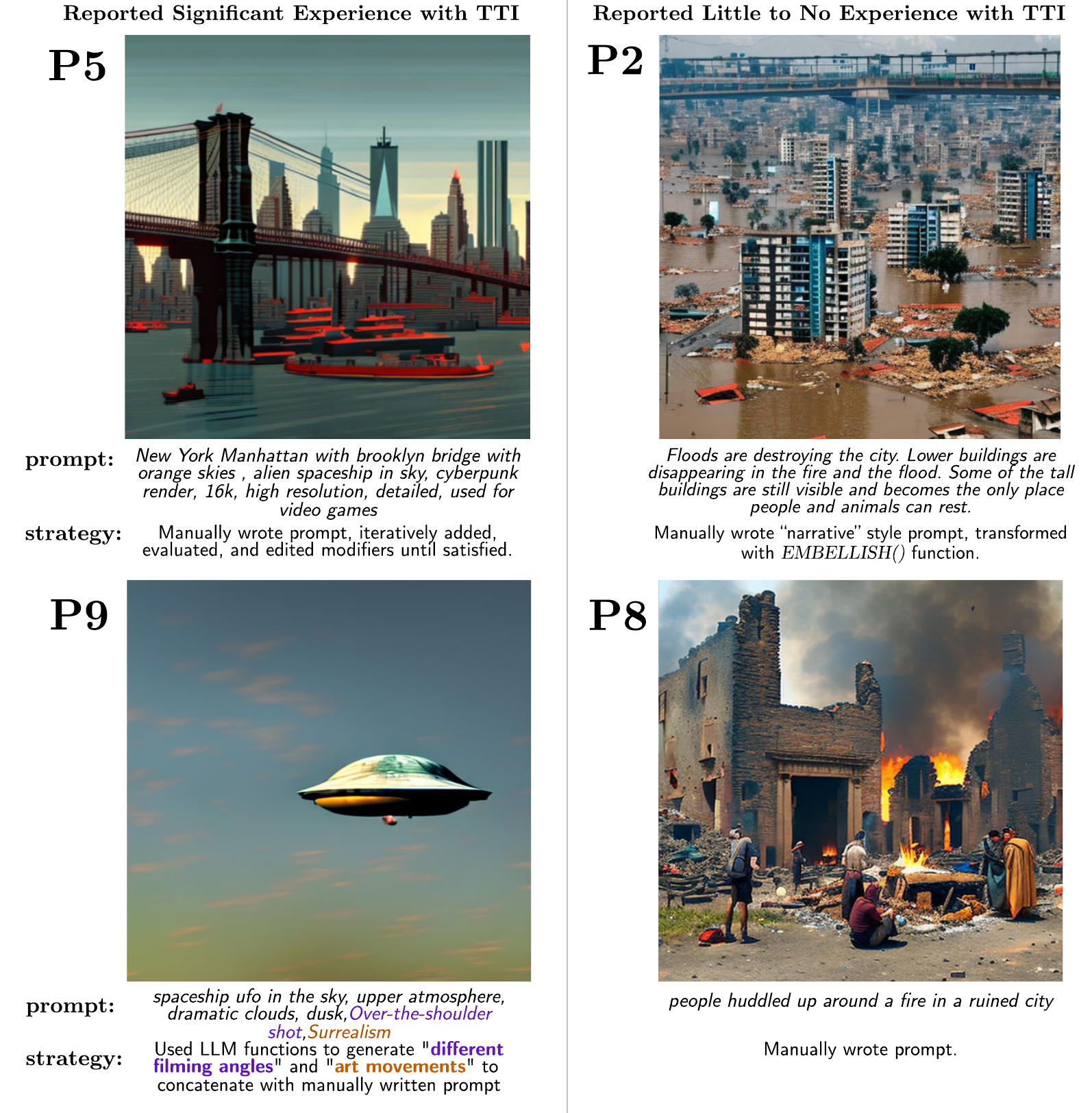}
    \caption{Examples of prompts authored by participants in the first study}
    \Description{Examples of prompts authored by participants in the first study. Left column provides examples of images and prompts authored by users with significant experiences with TTI while the right column provides examples for users who reported little to no TTI experience.}
    \label{fig:study1results}
\end{figure}
We discuss usage patterns and themes informed by \textit{both} studies in section~\ref{sec:findings}. 

More than half of the participants in this study (7 out of 10) reported limited to no experience with TTI systems, but \textit{all} participants were able to successfully utilize a prompt-crafting workflow in the \tool system, and to produce generations that they were satisfied with for the concept art task.
Though authored directly by participants, the workflows adopted by more novice participants were likely inspired by the example structures showcased during the initial tutorial phase, with P3 and P4 copying from the tutorial examples directly. 

The seven participants with limited prompt-engineering experience (P1, 2, 3, 4, 6, 8, 10) wrote their prompt in ``English'' ranging from brief sentence fragments to detailed scene descriptions. Meanwhile, 4 of the 5 participants who reported substantial or extensive TTI experience (P5, P7, P9, P11) wrote in a structure specific to ``prompt language''---comma separated lists of terms, including modifiers to influence visual style.

Novice and experienced participants found LLM-based functions useful for creating or improving their prompts.
Participants with more spreadsheet experience more readily adopted string concatenation strategies to construct prompts. Using LLM-based functions to generate a series of words in a particular category, like \textsc{gpt\_list}("camera angles") or \textsc{synonyms}("red"), participants introduced semantic modification ``slots'' into their prompts. We describe this \textit{LLM-assisted dynamic prompt construction} as a prompt-space exploration strategy in Section 5.3.

This formative study confirmed \tool's usefulness as an exploratory TTI system and illuminated promising usage patterns, as we observed even novice participants begin developing various strategies and structures to support their task completion goals in the \~40 minutes provided. However, the brevity and constraints (i.e. prescribed creative task) of this first-use study format meant that users did not fully leverage \tool to develop strategies for user-defined, creatively motivated goals. This motivated a longer-term expert study to observe how generative artists might utilize \tool to build systems for ``real-world'' creative workflows. 
\begin{figure}
    \centering
    \includegraphics[width=0.48\textwidth]{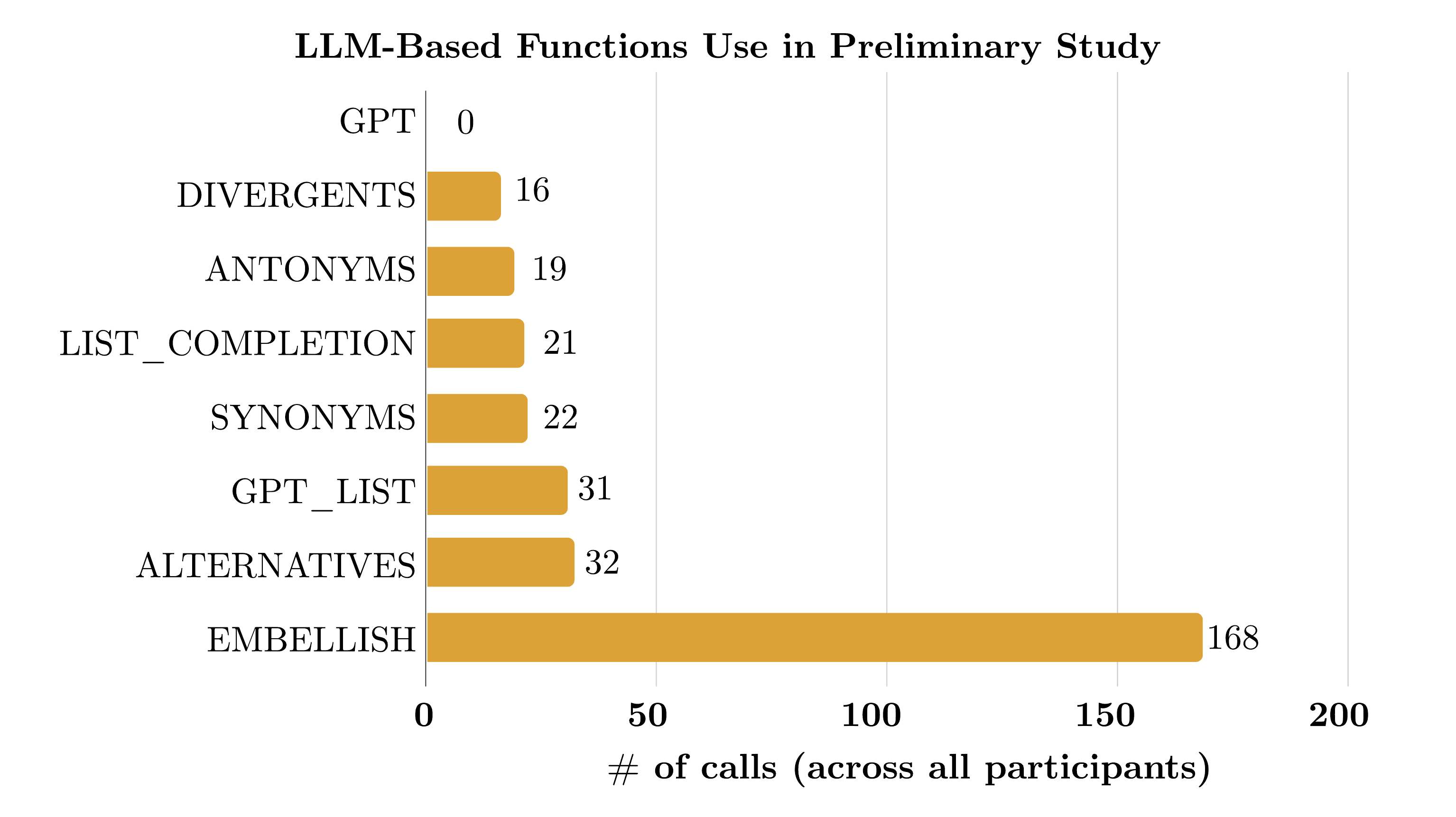}
    \caption{Number of times participants used each of the LLM-based functions (listed in order of ascending frequency) during the preliminary lab study activity. There were 4,737 calls to the TTI() image generation function.}
    \label{fig:frequency1}
    \Description{ Bar chart showing LLM functions on Y Axis against count of usage by participants on the X axis, from 0 to 200, increments by 50. 7 bars are shown. Embellish has the highest usage relative to the rest of the LLM functions.  }
\end{figure}
\begin{figure}
    \centering
    \includegraphics[width=0.48\textwidth]{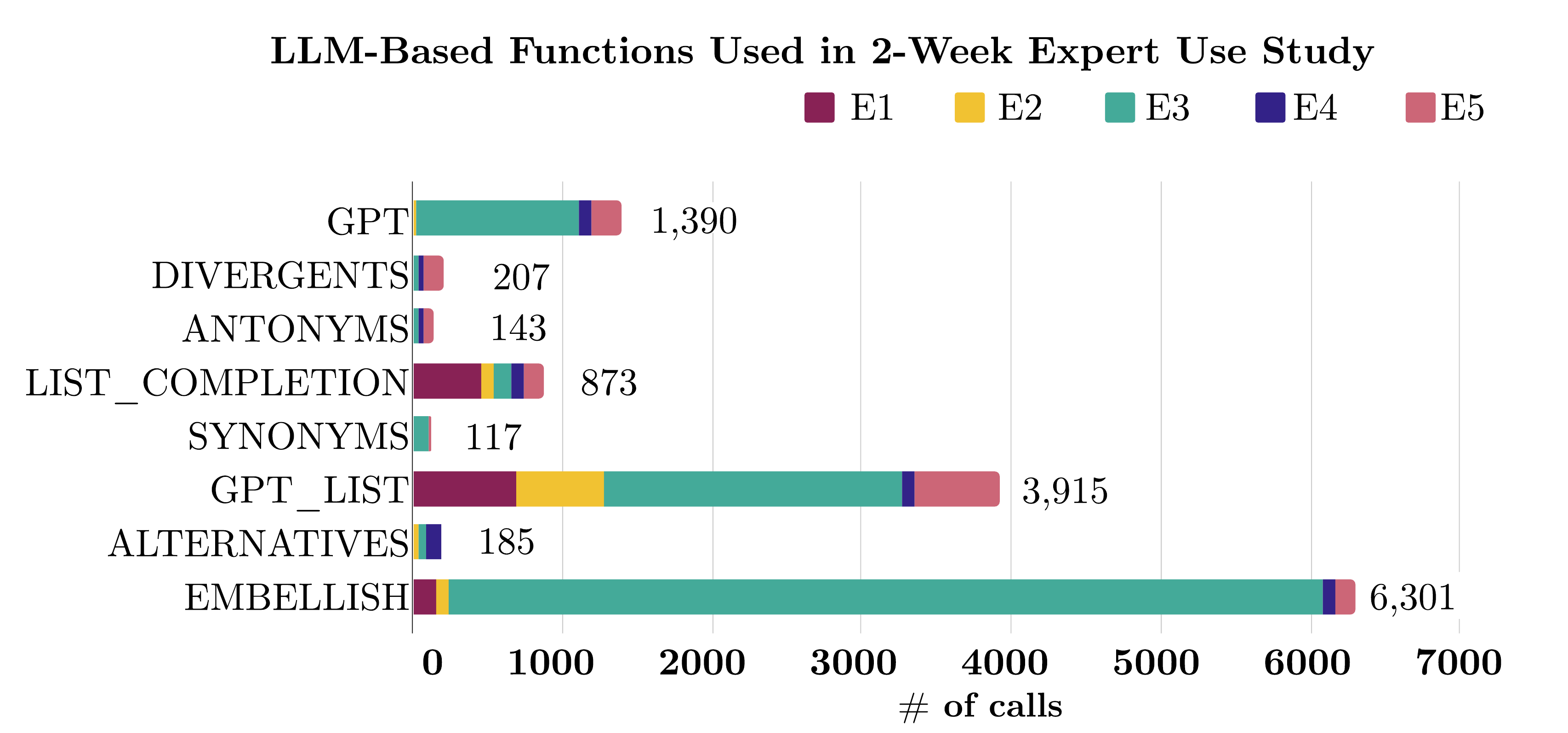}
    \includegraphics[width=0.48\textwidth]{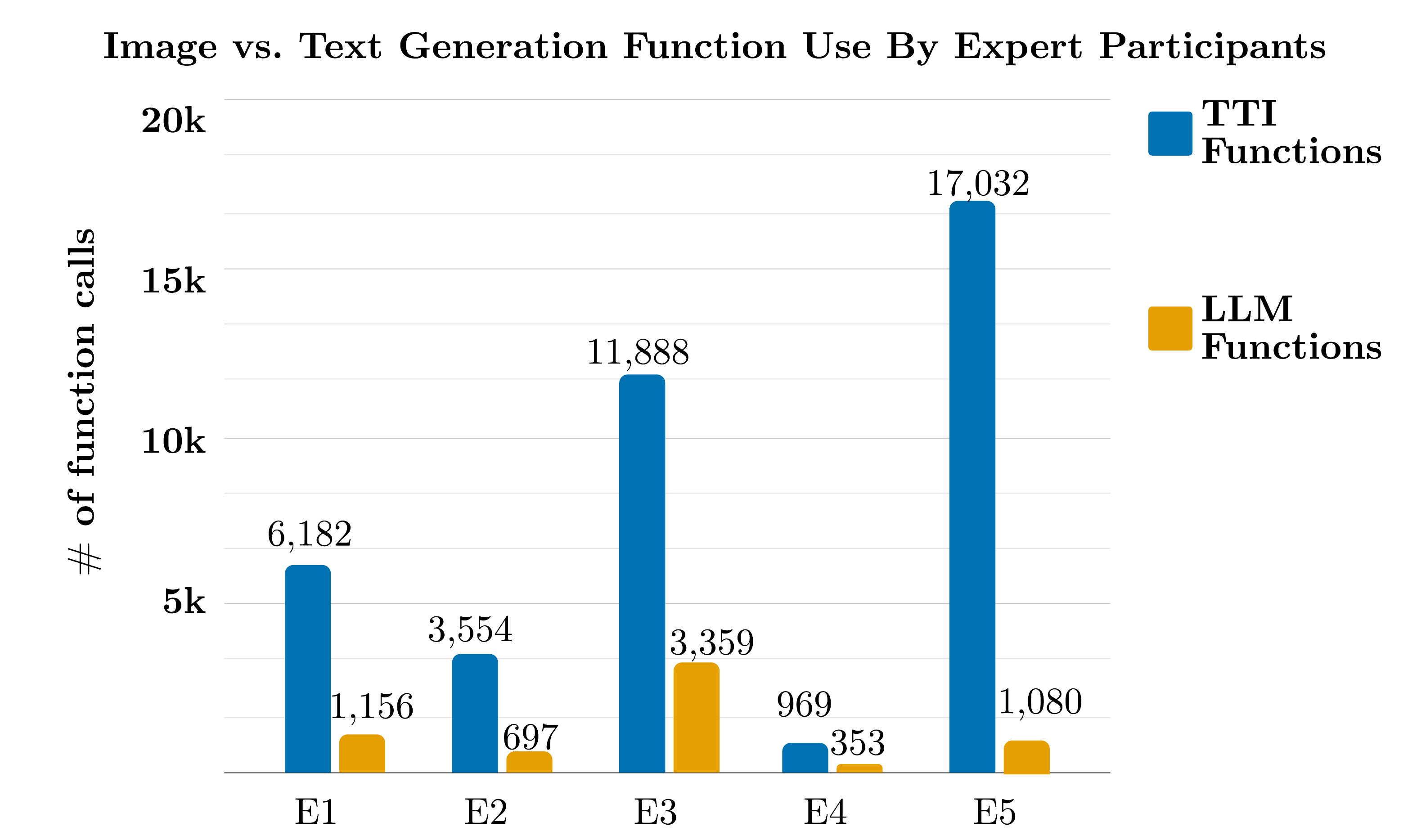}
    \caption{Above, LLM-based function use across the 2-week expert use study. Individual colors represent individual expert participants consistently across functions (5 total). Below, the number of times each expert participant made a unique call to the image generation and text generation functions during the 2-week study.}
    \label{fig:frequency2}
    \Description{Bar chart showing LLM functions on Y Axis against number of time  preliminary expert participants used each of them on the X axis, from 0 to 6000, increments by 2000. 7 bars are shown, and each bar is color coded, indicating the distribution of usage across each of the 5 expert participants. Embellish has the highest usage relative to the rest of the LLM functions, while Synonyms was the least popular function. Another Bar chart, total number of function calls on the Y axis against each preliminary expert participant on the X axis, split by the number of TTI function calls and LLM function calls. Participants all had a greater number of TTI function calls compared to LLM function calls. }
\end{figure}
\subsection{Extended Expert Study Design}

We turned to an extended study to observe the custom sheet-systems that experts would create
when given the time and flexibility to pursue authentic creative explorations.
\subsubsection{Participants}
To recruit experts for our second user study, we sent recruitment messages to individuals publicly participating in generative art communities on social media, and recruited 5 individuals (designated as E1-5 to differentiate from Study 1 participants).

\subsubsection{Protocol}


We conducted three 45 minute interviews spanning 2 weeks with each participant. 
Participants were instructed to use the tool for about 7-10 hours over the course of the 2-week study. We suggested 30-45 minutes of tool use per day, but participants were given the freedom to decide the length and structure of their work sessions.
As with the first study, the initial interview began with a short tutorial reviewing spreadsheet functionality and demonstrating \tool functions. 
The collaborative spreadsheet shared with each participant included documentation and examples of \tool function use. Participants could contact the research team via email with questions throughout the study. 

The second interview took place 1 week into the study. We asked participants to explain their exploration goals and strategies, and to use relevant parts of their spreadsheets to illustrate. We described back to participants our observations, allowing them to clarify any potential misinterpretations of their actions.
Based on the feedback we received, we designed a UI mockup that incorporated elements inspired by the structures built and functions used by participants during their first week of using \tool. 

In the third and final 45-minute interview, we again asked participants to describe the creative explorations evident in their spreadsheets, and to explain how they integrated \tool's functionalities into their creative process.
We then showed participants the UI mockup to gain their perspective and elicit further feedback and suggestions for designing more supportive TTI interfaces. 
\subsubsection{Data Collection and Analysis}
We analyzed our participants' usage of \tool as observed or described during interviews, as well as the resulting artifacts:  the sheets and usage logs, containing the full chronology of function calls made to the \tool system. 
We periodically viewed their spreadsheets throughout the 2-week study, including their Version History – a detailed record of changes made to the sheet, which we used to recover and save copies of previous versions.
\edits{We engaged in a rigorous data coding and analysis process by stepping through the version history of each expert participant's Google Sheets document and leveraging usage data logs to build data visualizations of each ``exploration session.'' This included labeling high-level screenshots of each sheet noting structures (e.g. exploratory axes like ``different seed in each column''), clustering semantically-similar prompts, and modeling the development of prompts over time, denoting both LLM-generated and manually authored iterative prompt modifications similar to those taxonomically defined in \cite{oppenlaender2023taxonomy}.}

\section{Findings: Structures for Exploring Prompt-Input-to-Image-Output Space}
\label{sec:findings}
Across both studies, participants' use of \tool and the artifacts they produced allowed us to observe and identify key elements of their TTI creative workflow: their \textit{goals}, the \textit{strategies} they chose to pursue, and the interface \textit{structures} they constructed to support these strategies. 


\edits{Our expert participants were able to construct a number of sophisticated spreadsheet systems and improved them throughout the study. The design of these systems were shaped by their particular} exploration strategies for navigating TTI space; \edits{ we highlight some of the patterns that emerged} in the findings below. 

\edits{In this section, we abstract the hyperdimensional Prompt-Input and Visual-Output space as two dimensions that the generative model maps between. The hyperparameter input space (2-dimensional in \tool, using \textit{seed} and classifier free guidance, or \textit{cfg}) is shown, when relevant, as a smaller, transformative dimension that lies between the two hyper-dimensions and influences the mappings between them. We use these abstractions to illustrate a particular exploration strategy and draw attention to the particular space(s) that TTI users are sensemaking during their exploration, and show structures prototyped by \tool participants that exemplify that exploration strategy.}

\edits{The overall trend was for participants to start by recreating the prompt templating activities as in other tools, but then build on top of these by selecting and combining axes of exploration–which can operate in both linear and non-linear ways. Participants then constructed 2D “small multiples”-style grids, iteratively layering axes for increasingly sophisticated explorations of prompt-image space---towards targeted areas of image space, and simultaneously, towards sense-making exploratory goals: to observe and understand capabilities and interactions. For example: what artists and styles can this model reproduce? What subjects (e.g., animals) and attributes (e.g., colors, facial expressions) might interact to yield interesting results?}
\edits{
\subsection{Iterative Prompt Exploration}}

\begin{figure*}
    \centering
    \includegraphics[width=0.95\linewidth]{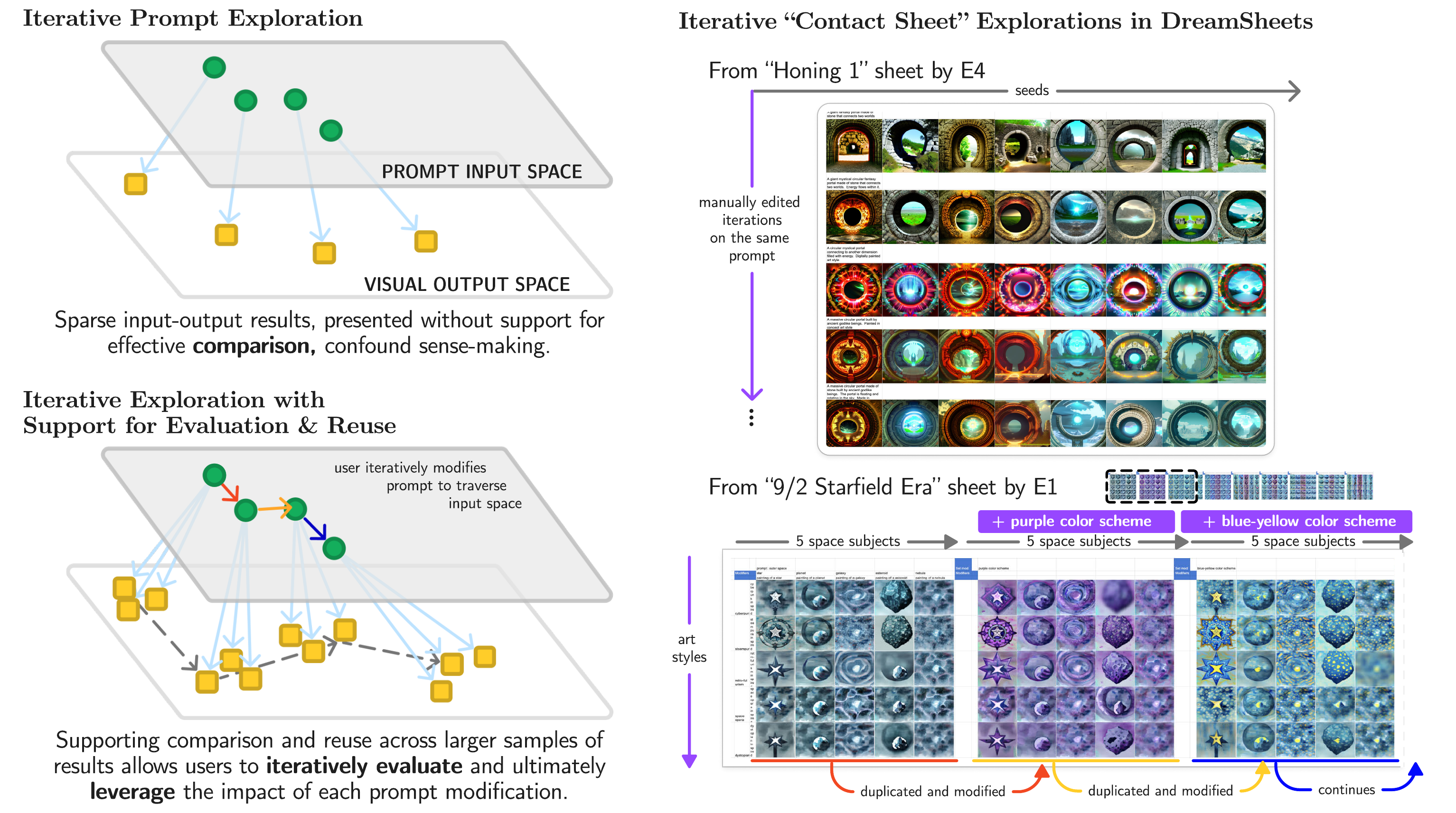}
    \caption{
    We abstract the Prompt-Input and Visual-Output space as two dimensions that the generative model maps between, and use these abstractions to illustrate the space(s) that TTI users are targeting via different exploration strategies. 
    \edits{Steering towards desirable results with iterative prompt refinement is a common strategy, but sparsely presented results can confound productive sense-making. 
    \tool users prototyped iterative sheet-systems that leveraged the ``infinite-canvas'' and rich history-keeping affordances of digital spreadsheets to conduct large-scale comparison across results, and to arrange and repurpose past explorations. With a structured, scalable results display, users can stumble upon interesting outliers without being confounded by them.}}
    \Description{On the left, a conceptual diagram generally depicting Iterative Prompt Exploration. At top, green circles are positioned on the top-most layer, which is labeled “prompt-input space” Light blue arrows point to yellow squares on the bottom-most layer, visual output space. It is labeled, sparse input-output results, presented without support for effective comparison, confound sense-making. Below, a similar diagram, labeled Iterative Exploration with Evaluation \& Reuse Support but colored arrows direct from one green prompt point to the next on the top layer, labeled user iteratively modifies prompt to traverse input space.  3 light blue arrows per each green point map to multiple yellow squares on the bottom-most layer, visual output space, and dotted gray arrows direct from each output group to the next. It is labeled, Supporting comparison and reuse across larger samples of results allows users to evaluate and iteratively leverage the impact of each prompt modification. On the right, zoomed out screenshots of two participant sheets: Above, the sheet,  Honing 1 by E4 , which different seeds horizontally creating variation across several rows, each generated with manually edited iterations on the same prompt. Below, a small section of From “9/2 Starfield Era” sheet by E1 which shows several grids of images, each has the same 5 space subjects across the columns horizontally, and different art styles vertically. Colored arrows indicate how each subsequent grid is duplicated and modified from the previous. The second in the series added “purple color scheme” and the 3rd added “blue yellow color scheme” leading to similar series of generations with different colors.}
    \label{fig:iterative}
\end{figure*}

\edits{Iterative prompt refinement, where participants gradually refine a prompt while testing the effects of each addition, is a fundamental TTI exploration strategy possible in any interface. \shm{TODO: cite another paper that observed users iterating on a prompt? or maybe just move this stuff to background/related work } However, many interfaces provide only a few ($<10$) results, and offer limited support for comparing results, displaying only the results of one experiment at a time, or only allowing users visually compare the results of chronologically adjacent experiments, as in the \textit{sequential ``chat'' history} interface model used by Midjourney or Dreamstudio.

Limited support for user-structured comparisons makes it difficult to evaluate the impact of prompt modifiers, and confounds effective sense-making; what the user may perceive as a \textit{semantically close} edit in prompt space can translate to a confoundingly \textit{large visual transformation}, and vice versa -- see Figure~\ref{fig:iterative}.}

Our expert participants described previously using external tools to save and evaluate TTI results history -- saving favorite results and prompt modifiers into a spreadsheet (E1, E5) or word document (E2, E4) with notes on the expected impact of each token, for example. 

Spreadsheets \textit{inherently} afford rich, reconfigurable, and structured history-keeping and results evaluation. \tool users leveraged the``infinite canvas'' qualities of digital spreadsheets to keep and evaluate in situ records of their exploration history. They also leveraged the inherent \textit{reconfigurability} of results within \tool; all participants used duplication (copying and pasting groups of cells, or entire sheets) to repurpose and iterate on prior explorations (including novices, who duplicated and adapted liberally from the tutorial structures.)

\edits{To effectively steer prompts towards desirable outputs, users benefit from reconfigurable history structures. Concurrently, \tool users found that it was beneficial to \textit{generate} and \textit{view} larger samples of results simultaneously.} All participants tried organizing generated outputs in a ``small multiples'' or ``contact sheet'' layout (as described by E1); 3 of the 5 experts (E1, E4, E5) explicitly remarked on its usefulness for large-scale results evaluation. Calls to the \textsc{tti} function comprised the bulk of participants use of \tool, as shown in Figure~\ref{fig:frequency2}; generating an average \textbf{7,925 unique images }over the course of the study. \shm{so many credits lol}

\edits{To \textit{generate} these larger samples of results, participants used \textit{variation} strategies to efficiently and systematically generate many variations of a single prompt idea. These variation strategies took the form of \textbf{Parametric Manipulations} and \textbf{Semantic Manipulations} as described in the following subsections.}

\edits{\subsection{Parametric Manipulation}}

While prompt-crafting is central to the effective use of TTI models, the ability to quickly manipulate hyperparameters alone provides a useful dimension for exploration, 
\edits{ this motivated participants to develop structures around hyperparameter control.

All expert participants used dynamic references to a column or row with a series of hyperparameter values to prototype a ``slider'' like evaluation structure; see Figure~\ref{fig:parametric_linear} for examples using \textit{cfg}. 

3/5 expert participants (E2, E3, E5), used ``Power Cells'' to prototype the functionality of a global ``Settings'' panel with the option to \textit{regenerate on update}. By structuring sheets such that all generations reference a seed or cfg value from a particular cell, updating this ``Power Cell'' would regenerate the entire sheet of results. This afforded iteratively testing values on a large sample of results to find a desirable ``setting.'' See E5’s “Same Seed Prompt Explore” in Figure~\ref{fig:parametric_stochastic}B for an example. }

\edits{\tool provides two parameters that influence image generation: a stochastic hyperparameter (\textit{seed}) and a non-stochastic hyperparameter (classifier-free-guidance, or \textit{cfg}) which participants used to transform their explorations in different ways.
 }
\subsubsection{Stochastic Transformations}
Seeds define the specific random noise that the diffusion model will use as a starting point; the model then repeatedly ``de-noises'' successive versions to generate an image, with the text prompt as a guide. 

\begin{figure}
    \includegraphics[width=0.44\textwidth]{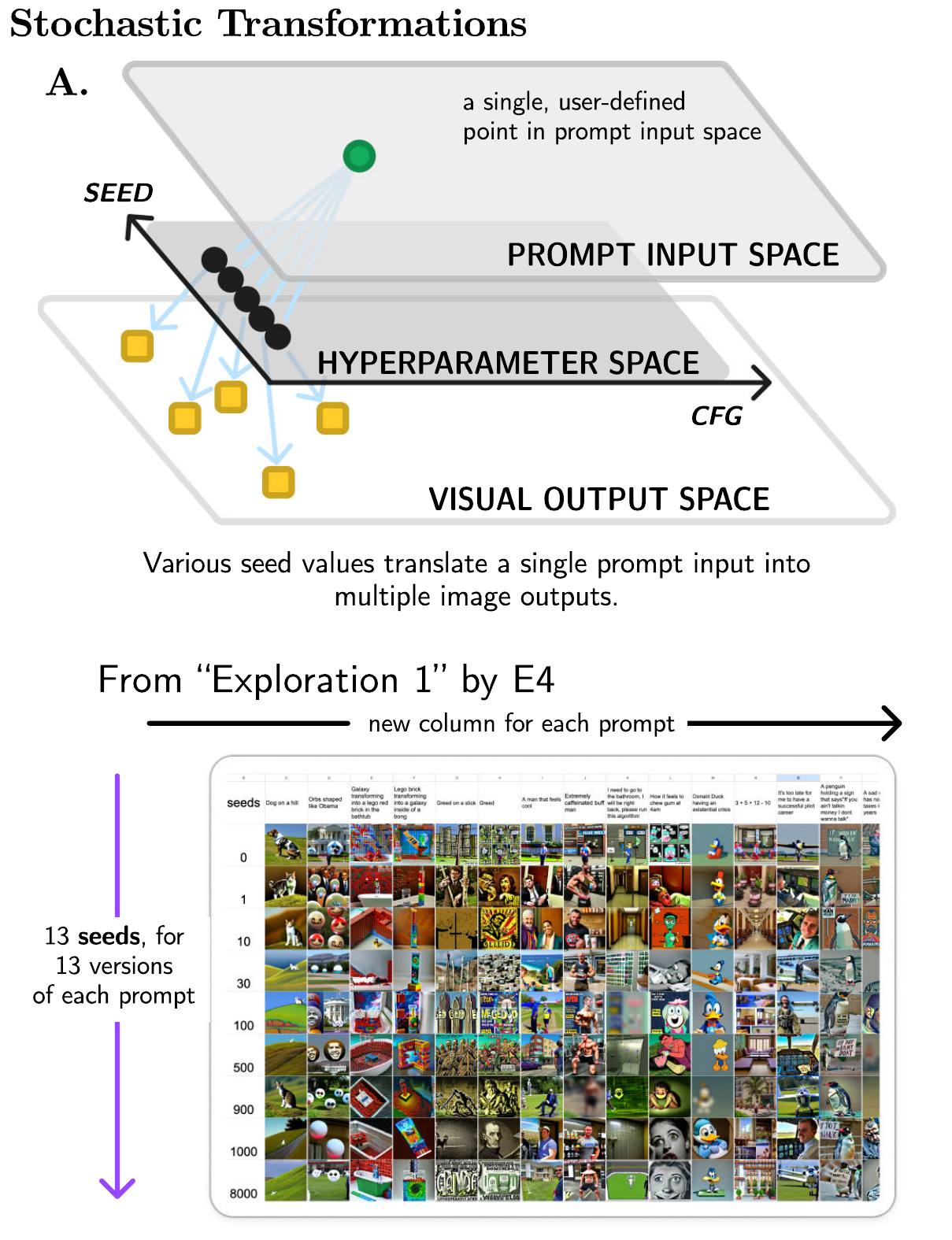}
    \includegraphics[width=0.44\textwidth]{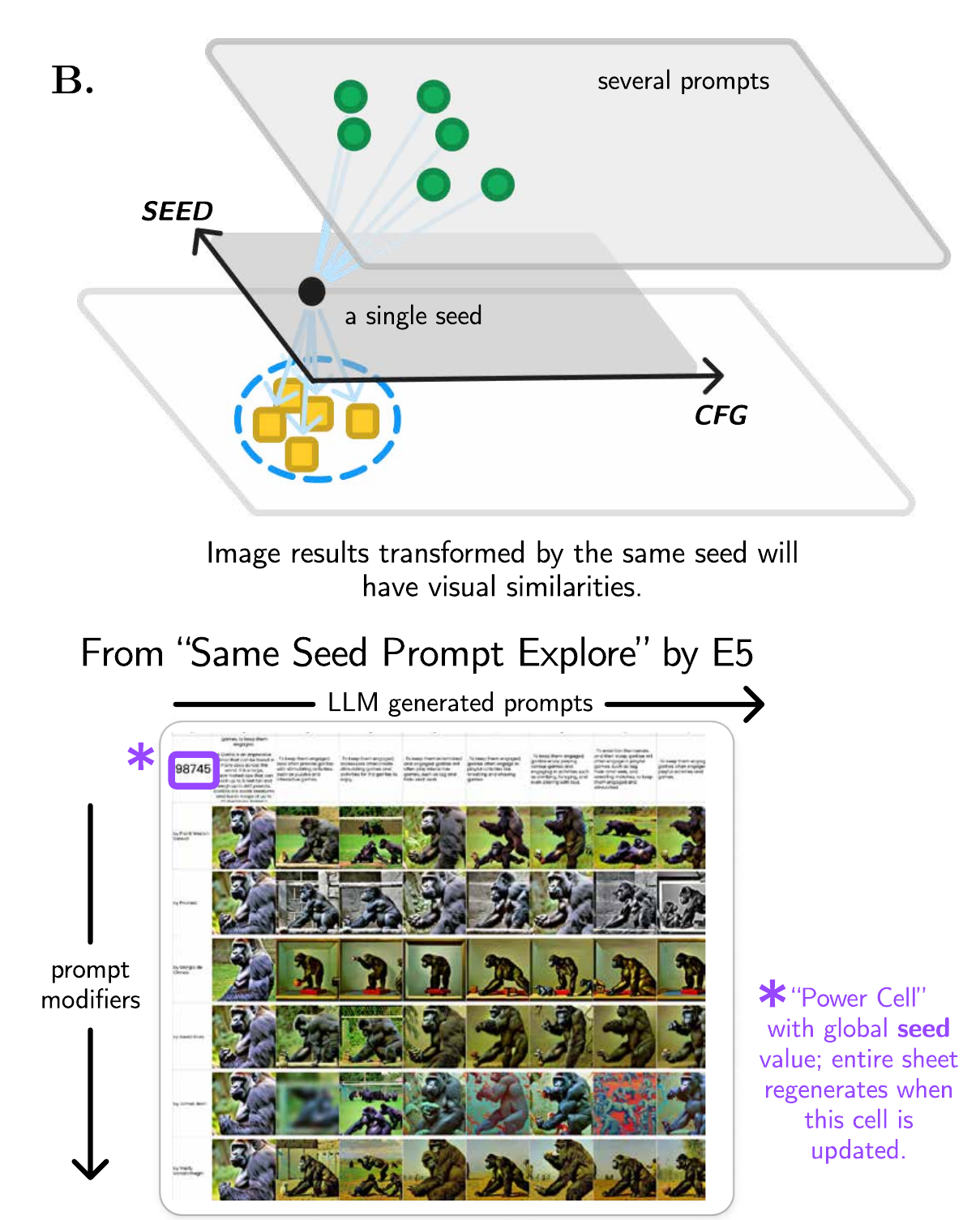}
    \caption{Here we introduce Hyperparameter space, a smaller intermediate dimension that can transform the mappings between Prompt-Input and Visual-Output space.
    Here we show two distinct strategies utilizing stochastic hyperparameters to transform results, with views of exemplary structures built by expert participants. A improves the efficiency of each test by increasing output sample size. B supports searching for and selecting a desireable hyperparameter value.}
    \label{fig:parametric_stochastic}
    \Description{At the top, a conceptual diagram labeled A depicting Stochastic Transformations. A green circle is positioned on the top-most layer, which is labeled ”a single user-defined point in prompt-input space.” Light blue lines show how this point maps to 6 black circles positioned on a middle layer, which has two axes indicated by labeled arrows: Seed and CFG, and is labeled HYPERPARAMETER SPACE. The black circles are positioned vertically, indicating that they are a linear series of seed values. Blue lines show how the input is mapped through this space to the bottom-most layer, visual output space, which has corresponding yellow squares placed randomly. Different seed values transform a prompt input, translating to several image results from a single input prompt. Directly below, a zoomed out screenshot of a participant’s sheet:  Exploration 1 by E4 , which has a new column for each prompt horizontally and 13 seeds with 13 version of each prompt vertically, indicated with arrows. Below this, a conceptual diagram labeled B, where several prompts (represented as green circles on the diagram’s topmost layer) mapping (with light blue lines) through a single seed (a black circle in hyperparameter space) to a group of several generations  on the bottom layer, indicated as yellow squares in a blue circle. Image results transformed by the same seed will have visual similarities. Directly below, Same Seed prompt Explore by E5 shows LLM generated prompts horizontally and prompt modifiers vertically, to generate a sheet of gorilla images. There is a purple box around a number in the top left corner of the sheet, which then is labeled “Power Cell with global seed value; entire sheet regenerates when this cell is updated.”}
\end{figure}

Generating many images using the same \textit{prompt} but different \textit{seeds} was a common strategy across participants in both studies. 

All 5 expert participants utilized seed variations to quickly evaluate many ``versions'' of the same prompt. \edits{As shown in Figure~\ref{fig:parametric_stochastic}A, a series of seed values allows the user to reveal a larger area of image-output space with each prompt. This improves efficiency towards creative goals (increasing the likelihood of finding a desirable output) and in sense-making (revealing larger samples of output space with each input test.) }

There is no perceptual correlation between adjacent seeds, but images generated with the same seed may share visual similarities with the original noise pattern. \edits{A seed can then be useful for biasing generations towards a particular composition or color pattern, motivating targeted explorations of hyperparameter space. E5 prototyped several versions of a seed-exploration structure, including "Same Seed Prompt Explore" shown in Figure~\ref{fig:parametric_stochastic}B.} With a ``vector graphics'' design goal, they used this structure to identify seeds that could bias the image generation to feature ``a central object'' on a flat background. E5 found seed \textbf{7935} and used this value in future ``vector graphics'' style explorations (see Figure~\ref{fig:multidimensional}B). 
\subsubsection{Non-Stochastic Parametric Transformations}
\begin{figure}
    \centering
    \includegraphics[width=0.45\textwidth]{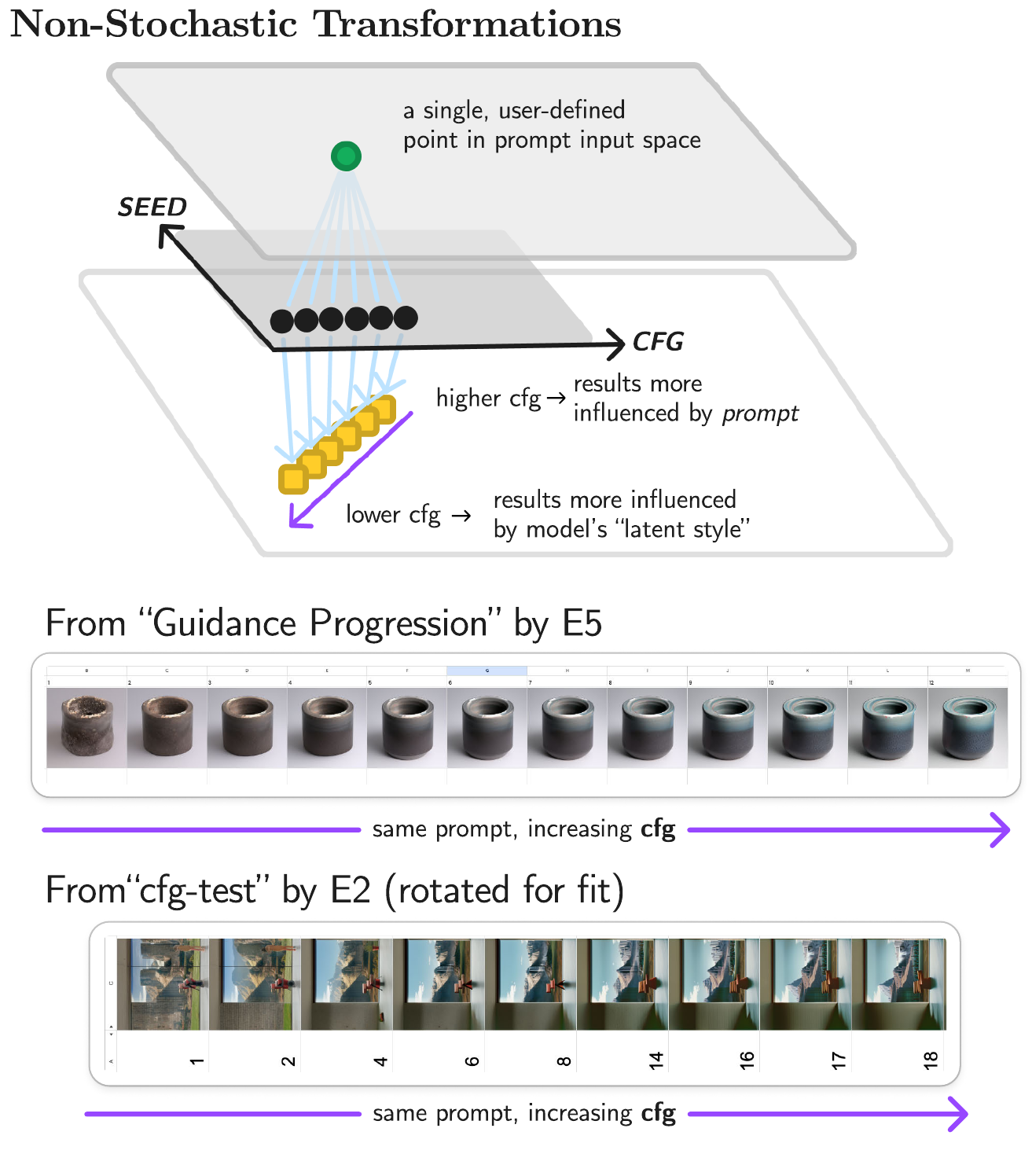}
    \caption{Participants created a ``slider'' along a series of classifier-free-guidance (cfg) values to preview a perceptually linear transformation on a particular generation. The global ``Power Cell'' structure shown in Figure~\ref{fig:parametric_stochastic}B was also used to iteratively test cfg values on a batch of results. Offering more``controlled'' transformation is particularly useful for ``depth'' explorations of image results, which DreamSheets otherwise lacks support for. }
    \label{fig:parametric_linear}
    \Description{At the top, a conceptual diagram depicting Non-Stochastic Transformations.. A green circle is positioned on the top-most layer, which is labeled ”a single user-defined point in prompt-input space.” Light blue lines show how this point maps to 6 black circles positioned on a middle layer, which has two axes indicated by labeled arrows: Seed and CFG. this indicates a 2 dimensional parametric transformation space. The black circles are positioned horizontally, indicating that they are a linear series of CFG values. Blue lines show how the input is mapped through this space to the bottom-most layer, visual output space, which has corresponding yellow squares placed along a purple arrow labeled “higher cfg = results more influenced by prompt, lower guidance value = results more influenced by model’s latent style.” Below, zoomed out screenshots of two participant sheets: guidance progression by E5 and cfg test by E2, rotated for fit. Both show a series of images that have a subtle linear progression in visual qualities, labeled with an arrow “same prompt, increasing cfg” pointing to the right.}
\end{figure}
E2 and E5 were particularly interested in exploring different \textit{cfg} values. 
\edits{ This hyperparameter has a perceptually linear influence on the image generation; a higher \textit{cfg} value generates images more strongly influenced by the prompt. Low \textit{cfg} values allowed E5 to gain a sense of ``the model's priors,'' using a phrase commonly used in machine learning to refer to bias or preexisting (``prior'') beliefs. E5, E4, and E1 alluded to the ``default style'' latent to a specific image generation model as being important for prompt-artists to learn. 

Non-stochastic hyperparameters that offer more ``controlled'' transformation are useful for exploration ``depth'' (i.e., repeated image refinement)---a capability that \tool lacks support for, and that systems like ControlNet\cite{zhang2023adding} cater to.}

\edits{
\subsection{Semantic Explorations}}
\label{dynamic}
\begin{quote}
    ``Does Stable Diffusion know the same artists I do?'' (P11)
\end{quote}
\begin{figure}
    \centering
    \includegraphics[width=0.9\linewidth]{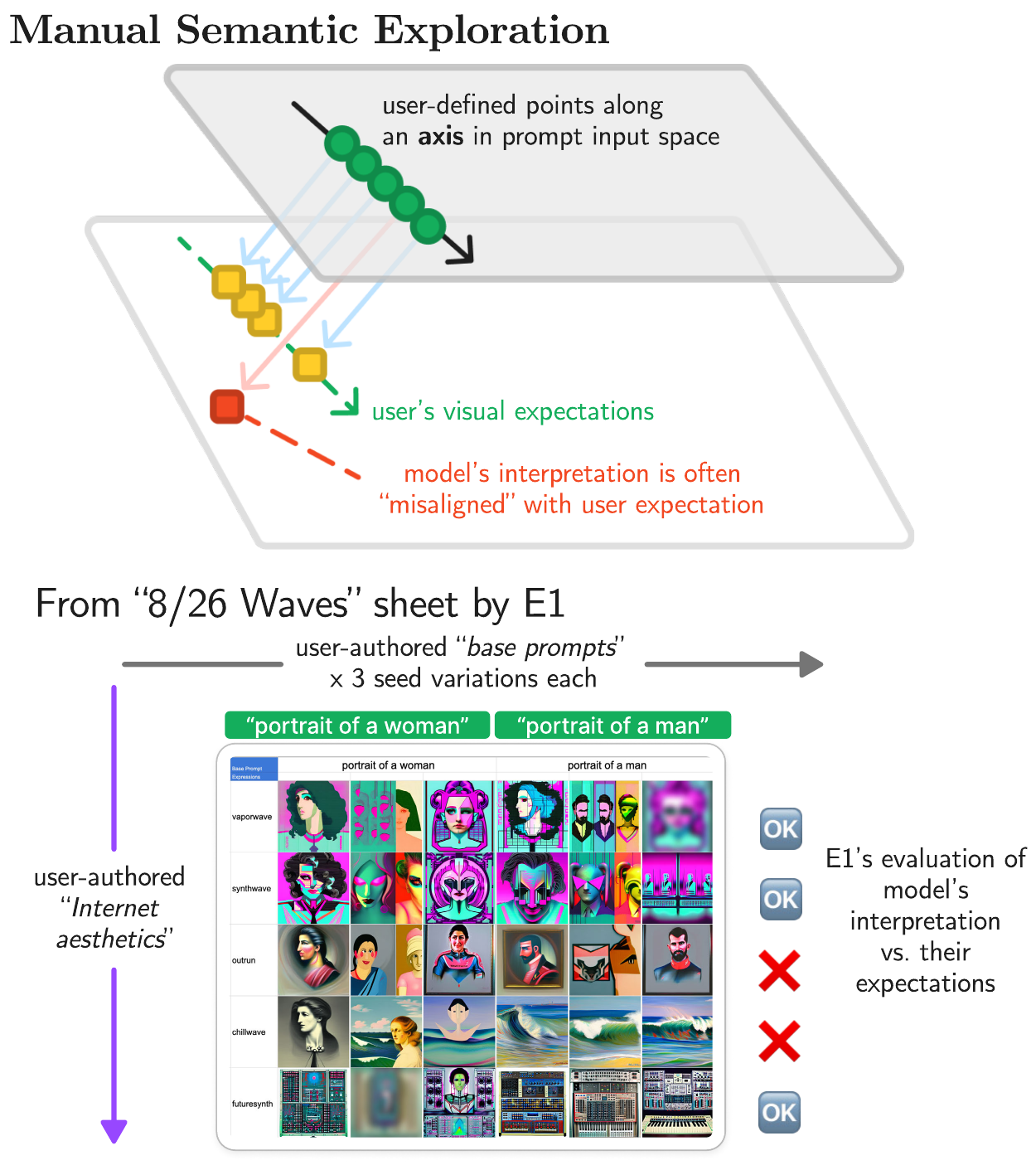}
    \caption{Participants constructed dynamic ``prompt templates'' that combine \textit{base prompts}
with ``swappable'' \textit{prompt modifiers} (here, a series of ``Internet aesthetics''). Users attempt conceptually gradual movements in prompt-space for procedural sense-making, iteratively testing how each modification to prompt text maps (often imperfectly) to movement in image-output space and updating expectations accordingly. }
    \label{fig:manualsemantic}
    \Description{At the top, a conceptual diagram depicting Manual Semantic Exploration. An arrow along which a small series of green points is positioned on the top-most layer, which is labeled ”user defined points along an axis in prompt-input space.” Light blue lines show how these points map to the bottom-most layer, visual output space, which has corresponding yellow squares placed along a dotted green arrow labeled “user’s visual expectations” indicating image generations. The fourth green point in prompt space, however, has a red arrow mapping to a red square in image space, which is labeled “model’s interpretation is often misaligned with user expectation.” Below, a zoomed out screenshot of a participant sheet; “From “8/26 Waves” sheet by E1. Horizontal axis arrow: user authored “base prompts” with 3 seed variations each, portrait of a woman and portrait of a man. Vertical axis arrow, user-authored “internet aesthetics”. The sheet depicts various women and men in colorful aesthetic styles, though some do not met the user’s expectation - the “chillwave” row has generated literal ocean waves. Emojis at right indicate E1’s evaluation of the models’ interpretation of each internet aesthetic against their expectations (two of the rows are misaligned with their expectation.)
    }
\end{figure}
Participants manipulated language to make movements in prompt-space that would, ideally, translate into movements towards interesting areas of image-space. The spreadsheet interface provided a familiar structure for 2D evaluations; evaluating the combinatorial effect of two exploratory ``axes'' at a time (e.g. ``subject'' columns by ``art-style'' rows) was a common strategy. P6, a novice user in the preliminary study with limited prompting experience, said:
\begin{quote}
    ``I think it's good to know how things change depending on different variables... the spreadsheet helps with navigating what exactly is changing within the image.'' (P6)
\end{quote}


\label{sec:axes}
\subsubsection{Manual Semantic Exploration}
\edits{To streamline iterative prompt explorations, participants constructed dynamic ``prompt templates'' that combine ``base prompts'' with swappable ``slots.'' Our expert participants echoed ~\cite{chang_prompt_2023}'s findings, treating these carefully crafted ``prompt templates'' as art pieces in themselves. 

In Figure~\ref{fig:manualsemantic}, E1 combined simple ``base prompts'' with several ``Internet aesthetic'' words to evaluate their efficacy as prompt modifiers -- checking if the model would interpret each ``aesthetic'' in alignment with their expectations. Participants used manually constructed series (as opposed to the list-generating LLM functions) to conduct specifically targeted explorations. E4 prioritizes total creative writing control, using TTI to craft comedic prompt-image pairs and narrative concept art; they chose to manually craft most prompts without LLM-assistance.}
\subsubsection{Generative Semantic Exploration}
\edits{Participants employed \tool's LLM-based functions and spreadsheet concatenation to build sheet-systems that streamline the discovery of useful points in prompt-space.
By crafting dynamic prompt templates} that reference from LLM-generated lists, participants select semantic ``axes'' to define a ``prompt space'' for LLM-assisted exploration.
4 of 5 experts (all except E4) utilized cell concatenation to craft dynamic prompt templates with LLM-generated prompt parts, though all of expert participants used the LLM functions to some extent. See Figures~\ref{fig:frequency1}~and~\ref{fig:frequency2} for counts of how frequently participants utilized each of the LLM functions across both studies. 

\begin{figure*}
    \centering
    \includegraphics[width=0.95\linewidth]{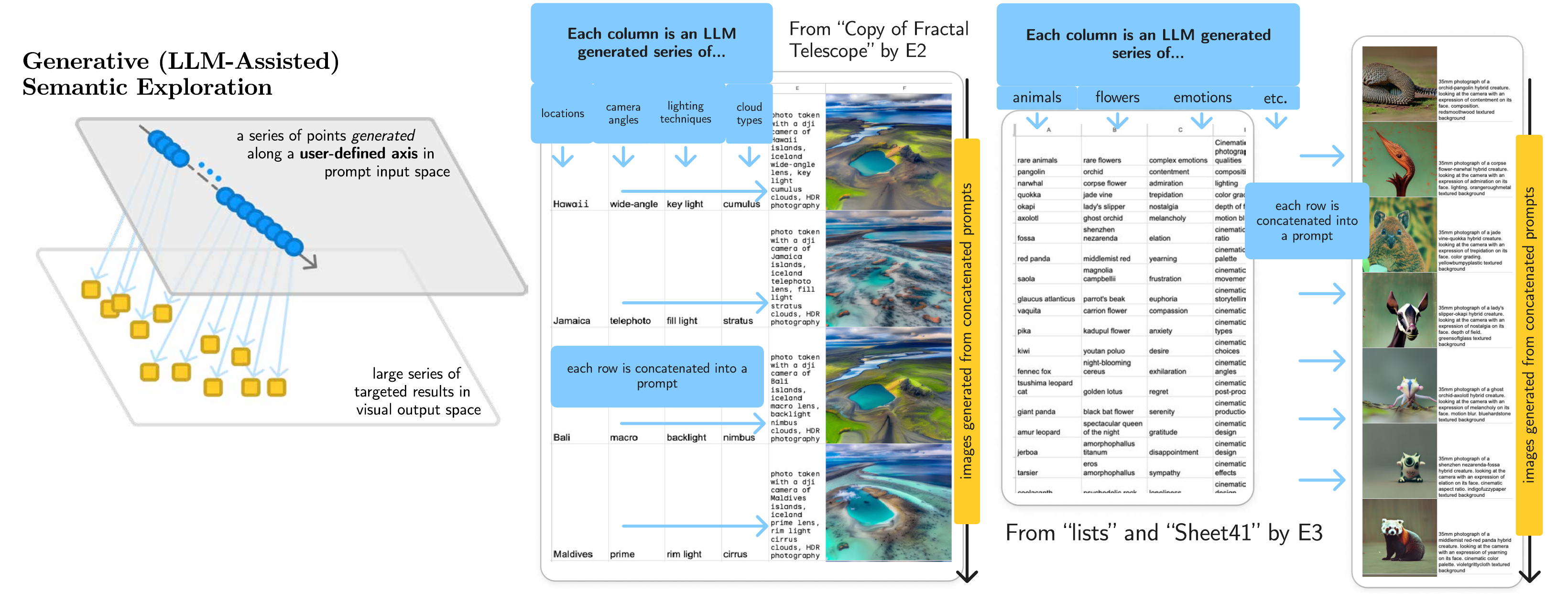}
    \caption{E2 and E3 independently developed conceptually similar sheet-system prototypes for different creative goals (drone-style landscape photography and animal-plant hybrids). They each used column-wise categorical-list generation (via LLM functions) and row-wise concatenation to generate a long column-wise series of results.}
    \label{fig:llmsemantic}
    \Description{On the left, a conceptual diagram depicting Generative (LLM-Assisted) Semantic Exploration. An arrow along which a series of blue points on the top-most layer, which is labeled ”a series of points generated along a user-defined axis in prompt-input space.” Light blue lines show how these points map to the bottom-most layer, visual output space, which has many yellow squares indicating image generations, and is labeled “large series of results in visual output space.” On the right, zoomed out screenshots of participants’ sheets: Copy of Fractal Telescope by E2 and “lists” and “Sheet 41” by E3. They both show that each column is an LLM generated series of camera angles, lightning techniques, animals, flowers, emotions, etc. Each row is concatenated into a prompt. There is a column of images with adjacent text, labeled “images generated from concatenated prompts”}
\end{figure*}
\edits{
E2 and E3 independently developed two-part interfaces that separate the design process into prompt-authoring and image-evaluation steps. The ``lists'' section of the interface houses the semantic-axes selection and sampling process: here, the user selects categories for text generation like ``lighting techniques,''  or ``mythical creatures,'' and decides how to combine them (often with row-wise, comma-separated concatenation.) 

In a separate section, the concatenated prompts are used to generate a column of image results. After constructing these sheet-prototypes, E2 and E3 reused them extensively for many explorations. 
}

For 3 of 5 expert participants (E1, E2, E5), list generation functions \textsc{gpt\_list} and \textsc{list\_completion} comprised more than half of their LLM Function use.  E3 also used these list generation functions many times---in fact, more than any other participant---but their use of \tool stood out overall: E3 made 9268 LLM function calls, 7.6 times more than the next most frequent user of the LLM functions.\deletes{; including 5851 \textsc{embellish} calls. See Figure ~\ref{fig:p3exploration} for a sample of the sophisticated text-manipulation workflow they developed.}

E1 used text-generation to discover interesting new prompt modifiers, including ``monochromatic,'' ``watercolor'', and ``pixel art.''
\begin{quote}
    ``I didn't set out to make pixel art or watercolors, but through the course of the study I discovered these aesthetic spaces that I really loved!'' (E1)
\end{quote}

LLM-functions can accelerate prompt space traversal while supporting creative \textit{Recognition over Recall}- users can choose \textit{camera angle} as a ``slider'' to explore, then recognize an appealing ``setting'' in the generated results - without having to know or recall the words to describe it ~\cite{Nielsen2022}.  

\edits{\subsection{Flexible Scaffolding for User-Structured Multidimensional Explorations}

Users approach TTI generation systems with a wide variety of creative goals, as showcased by the diverse images generated by our participants. 
Participants iteratively combined multiple, multidimensional exploration strategies to prototype bespoke sheet-systems for targeted explorations of prompt-image space. 


\begin{figure}
    \centering
    \includegraphics[width=0.48\textwidth]{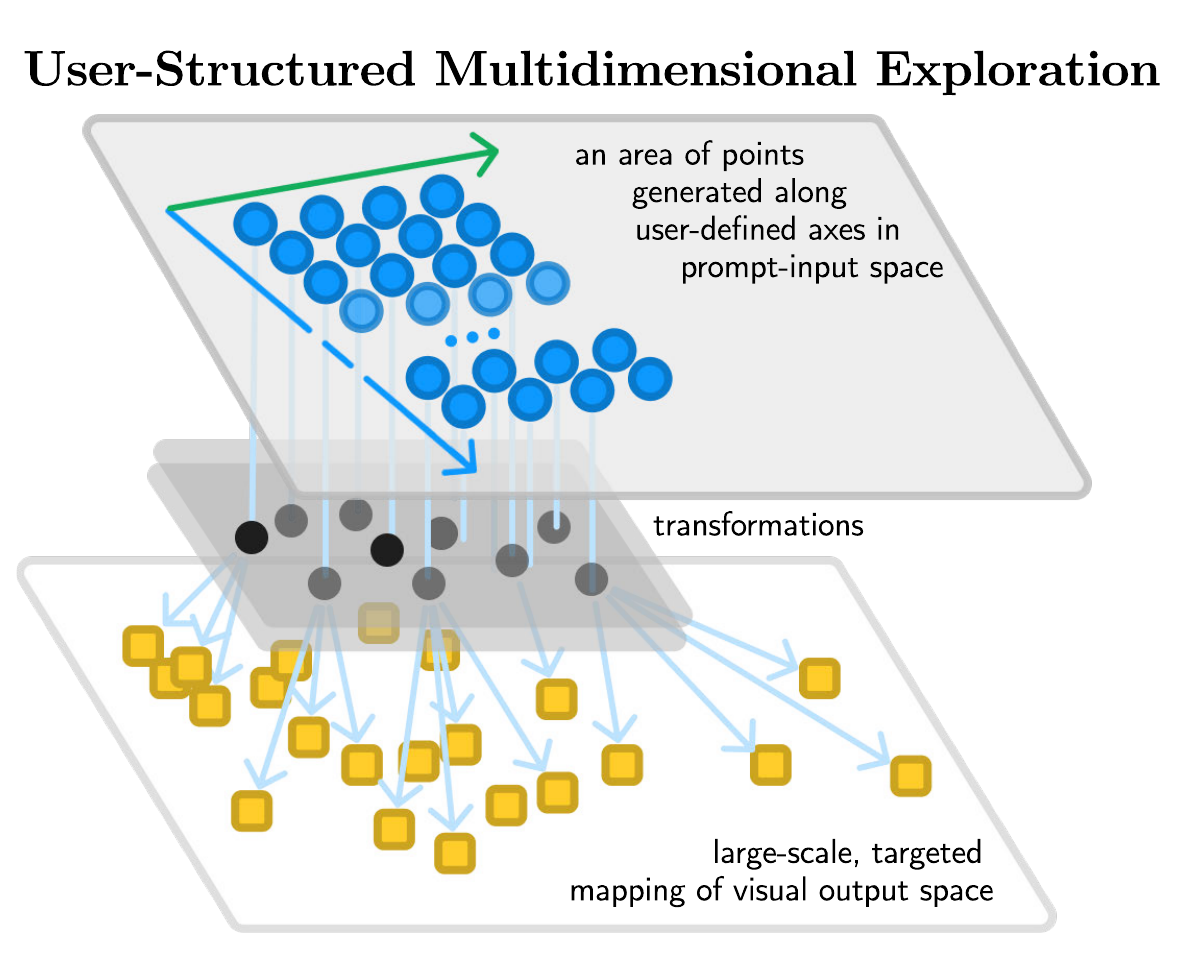}
    \\
    \includegraphics[width=0.48\textwidth]{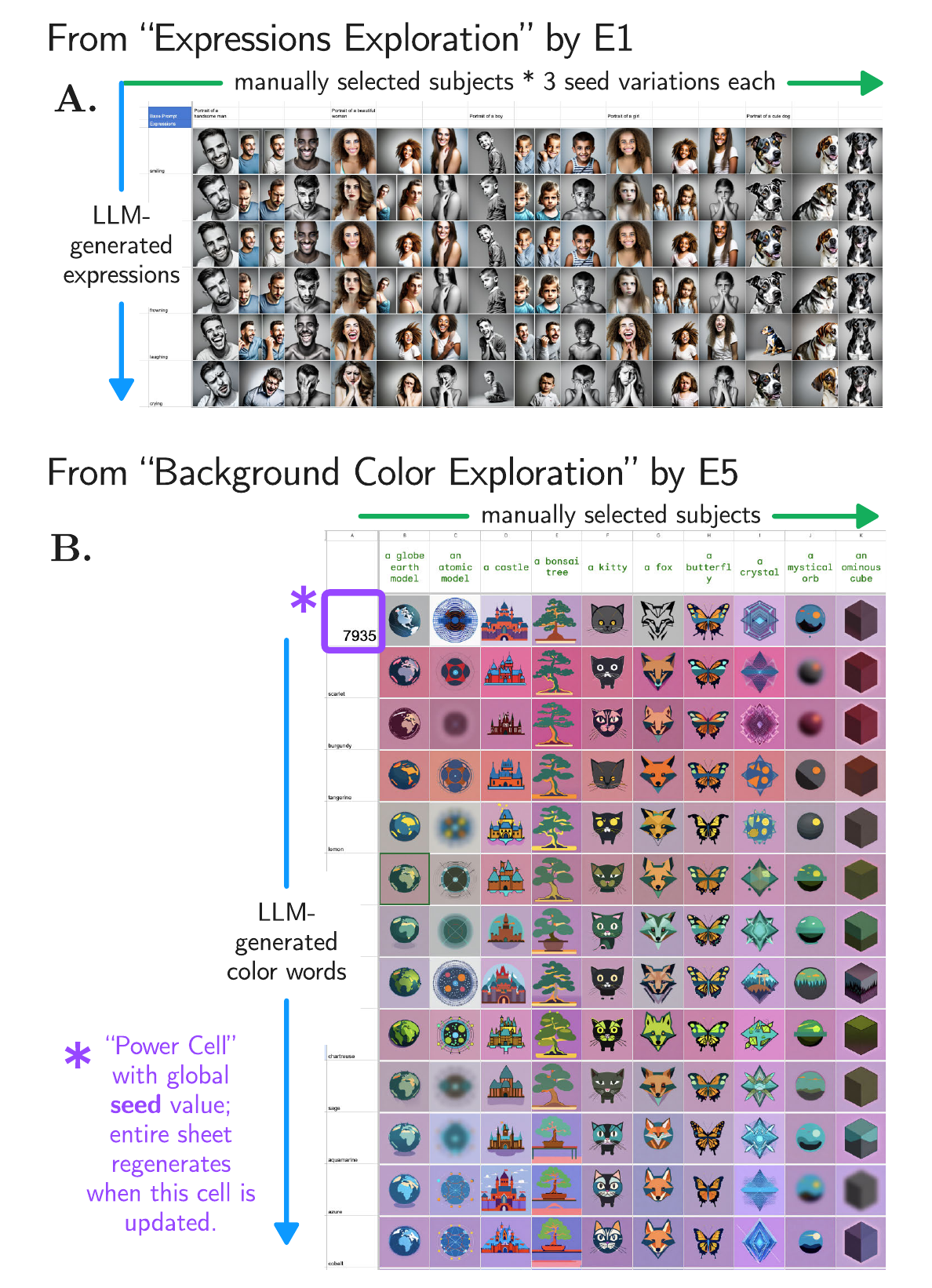}
    \caption{Participants adapted sheet-systems to combine multiple strategies to embark on a structured, multidimensional exploration, broadly evaluating a wide, but targeted, mapping of input-output space.}
    \label{fig:multidimensional}
    \Description{At the top, a conceptual diagram depicting User-Structured Multidimensional Exploration. Green and blue arrows define a 2D array of blue points on the top-most layer, which is labeled ”an area of points generated along user-defined axes in prompt-input space.” Light blue lines show how these points map through “transformation” layers in between prompt-input and the bottom-most layer, visual output space, which has many yellow squares indicating image generations, and is labeled “large-scale, targeted   mapping of visual output space.” Below, zoomed out screenshots of participants’ sheets:  First, “Expressions Exploration” by E1, which uses manually selected subjects with 3 seed variations each as the column-wise exploratory x-axis, and LLM generated facial expressions as a row-wise y-axis, to create a large-scale exploration of different expression words. Below this, a clipped screenshot of “Background Color Exploration” by E5, which shows many vector graphics style logo generations. E5 uses a series of manually selected subjects, a series of LLM generated color words, and Power Cell with a particular seed value that regenerates the sheet on update as a multidimensional combined exploration strategy to generate this distinct vector graphics art style. }
\end{figure}

\edits{Figure~\ref{fig:multidimensional}A shows some of E1's \textit{Expressions Exploration}} combining an LLM-generated list of \textit{facial expressions}, stochastic transformations, and a manually authored list of \textit{subjects} (man, girl, dog).

Figure~\ref{fig:multidimensional}B shows a segment of E5’s targeted, multidimensional ``vector-graphics’’ exploration, combining manual and LLM-generated semantic axes with a global ``Power Cell’’ for iteratively regenerating the sheet with different``flat-color-biased’’ seed values, identified via targeted explorations as in Figure ~\ref{fig:parametric_stochastic}B. 

 Over the course of the study, E5 generated 4693 such ``vector-graphics'' style images across two large-scale exploration sheets; E3 made 11,562 unique ``animal-plant photography'' generations with the system shown in Figure~\ref{fig:llmsemantic}.  Other participants pursued a variety of diverse creative focuses throughout the study, but all of these prompt artists exemplify the visual art style development observed by Chang \textit{et al}.~\cite{chang_prompt_2023}: with delicately crafted prompt-templates and deliberately selected hyperparameters, they developed distinct art styles and share with their online communities as images, prompts, and prompt-templates.

 \tool's flexibility allowed users to develop custom systems for various goals.  The patterns that nevertheless emerged suggest generalizable structures that future systems can offer as dimensional exploration ``units'' -- \textit{composable support structures} with the flexibility to decide when and how to combine them. This echoes a takeaway from Li \textit{et al.} in ``Beyond the Artifact: Power as a Lens for Creativity Support Tools'':  creative practitioners are empowered when they can \textit{laterally compose} tools in an efficient workflow, or refuse tools and replace them with others~\cite{li2023csts}.}

Flexible options for AI-assistance can play a role in fluidly supporting different exploration strategies and styles.   
When users hand off creative-labor and control to the power of AI-generated serendipity, they should maintain the power to reclaim control at any time. This echoes tensions observed by Lawton \textit{et al.} in ``When is a Tool a Tool? User Perceptions of System Agency in Human–AI Co-Creative Drawing''~\cite{lawton2023tool}. Future studies should investigate how co-creative systems might allow users to flexibly control where and how AI ``assistance'' influences their workflow.
\section{Co-designing Exploration Supportive UI Features with the DreamSheets 2.0 Mockup}

\begin{figure*}
    \centering
    \includegraphics[width=0.48\textwidth]{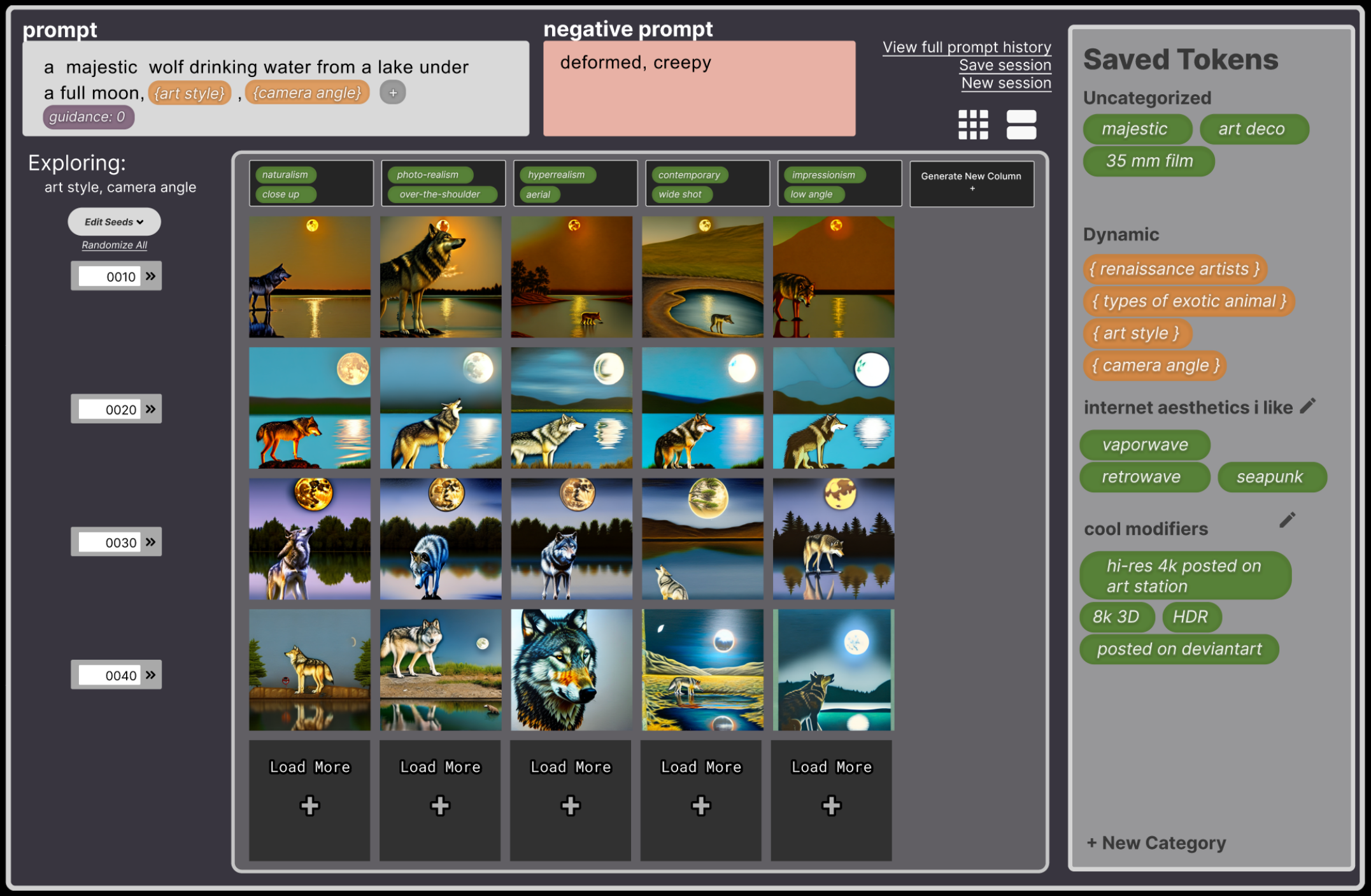}
    \includegraphics[width=0.48\textwidth]{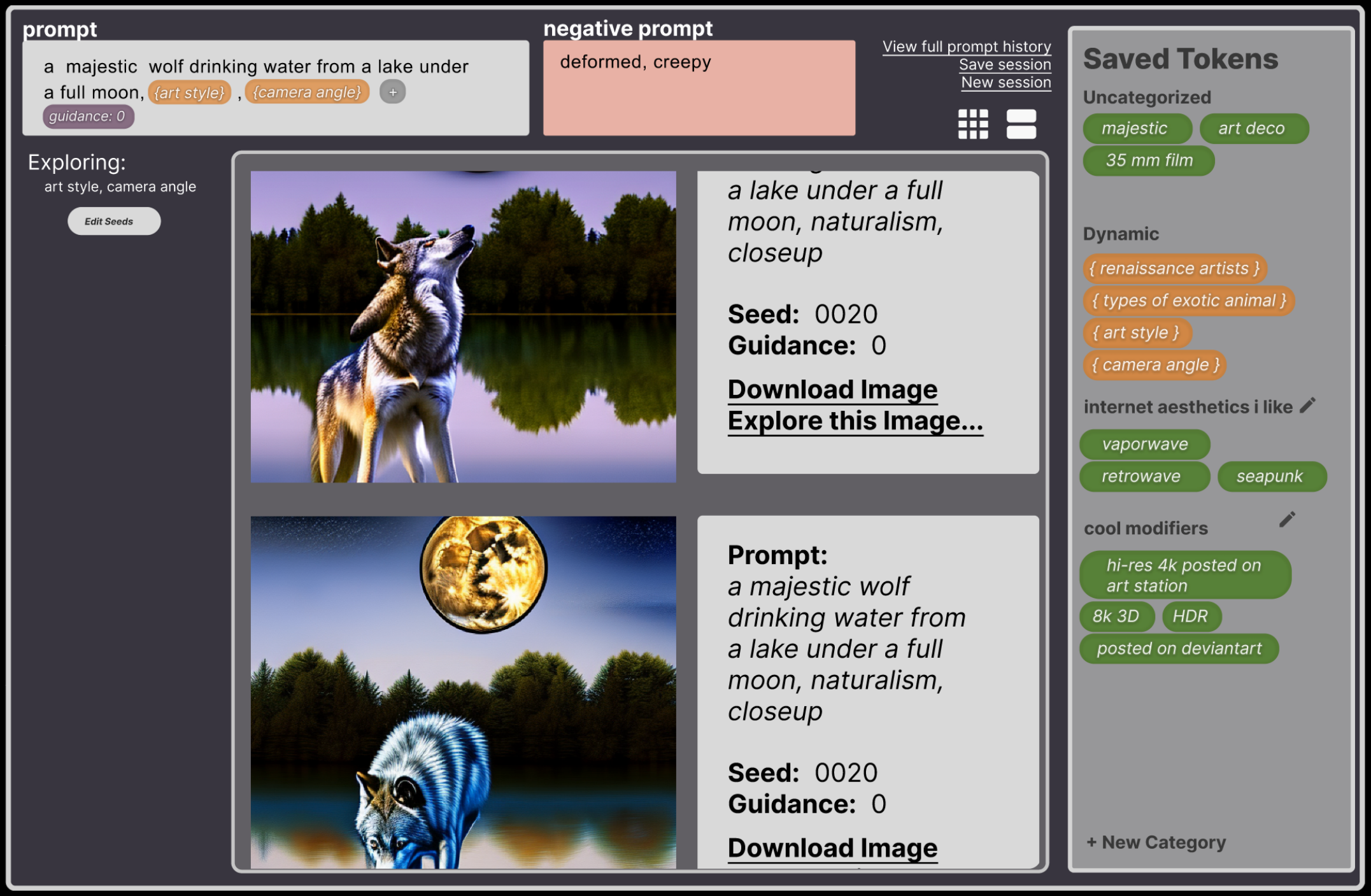}
    \caption{Views of the Dreamsheets 2.0 UI Mockup, showcasing the two visual layouts that users can freely toggle between: a grid view (left) and a focused list view (right). \tool 1.0 users valued the ``small multiples'' contact sheet layout, but viewing results in detail required them to change ``zoom'' settings or tediously adjust the sizes of columns and rows.  An improved UI could provide ``scalable'' displays to support TTI users as they move between broad and focused evaluations of results.}
    \Description{Views of the Dreamsheets 2.0 UI Mockup, showcasing the two visual layouts that users can freely toggle between: a grid view (left) and a focused list view (right). There is a Saved Tokens prompt bank as a right sidebar in both views. In the top left, a field for entering a prompt, with a red field for entering a negative prompt to its immediate right. There are options for changing the seeds used to generate each row of images.}
    \label{fig:mockuplarge}
\end{figure*}
\begin{figure}
    \centering
    \includegraphics[width=0.48\linewidth]{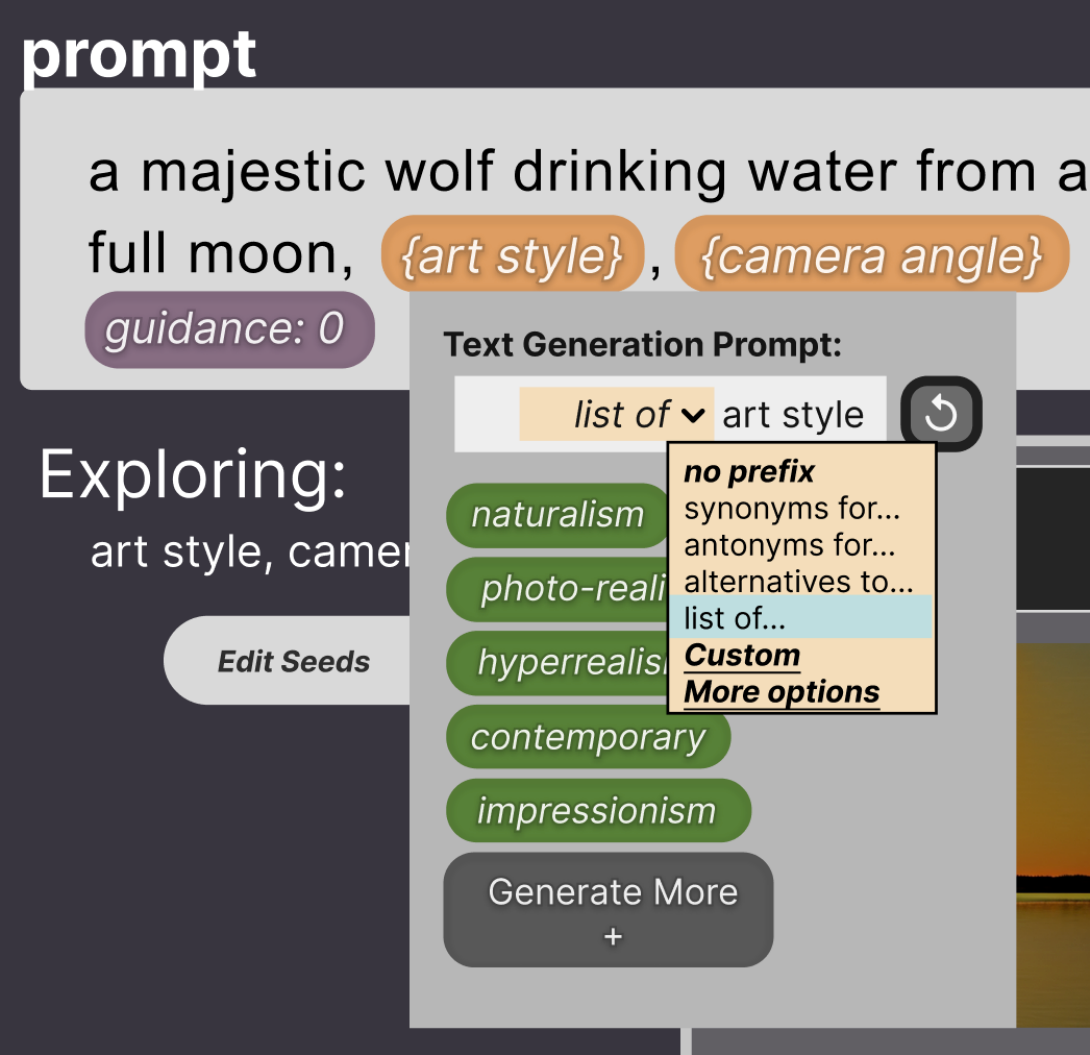}
    \includegraphics[width=0.46\linewidth]{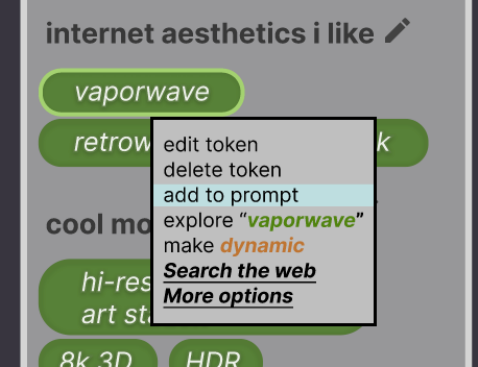}
    \caption{Views of elements featured in the interactive UI Mockup, including support for prompt templates with \textit{dynamic} tokens, allowing participants to flexibly generate semantic variations while eliminating the need to write cell concatenation formulas. The LLM supporting this interaction is surfaced; users can control how the prompt is sent.\\
    Tokens can be saved for future reuse.\\}
    \Description{Views of elements featured in the interactive UI Mockup. The left shows the “Dynamic Token” LLM menu, featuring options to change the prompt generating variations on “art style” The right screenshot shows a view of the prompt token box, including the interactive window that appears when the user clicks on a particular token, “vaporwave”. It showcases options to edit the token, delete, add to prompt, explore the “vaporwave” concept, make it into a dynamic token, or search the web.}
    \label{fig:mockupsmall}
\end{figure}

To more concretely probe into how participants conceptualized their own sense-making systems, and to inform a more generalized understanding of how we can support these processes, we developed ``\tool 2.0'' UI mockups with participants to elicit concrete feedback and speculative design ideas. 

To facilitate the application of our findings towards future interface designs, we map the co-designed UI concepts and components to the exploration strategies observed in participants' use of \tool 1.0; we present these in Figure~\ref{fig:uicomponents}.

\edits{\subsection{Rich, Reusable Exploration History}
Our participants' sheet-systems in \tool exemplified how TTI interfaces can offer improved support for iterative prompt exploration by affording rich history-keeping and structured, large-scale evaluation across results.
\begin{figure*}
    \centering
    \includegraphics[width=0.97\linewidth]{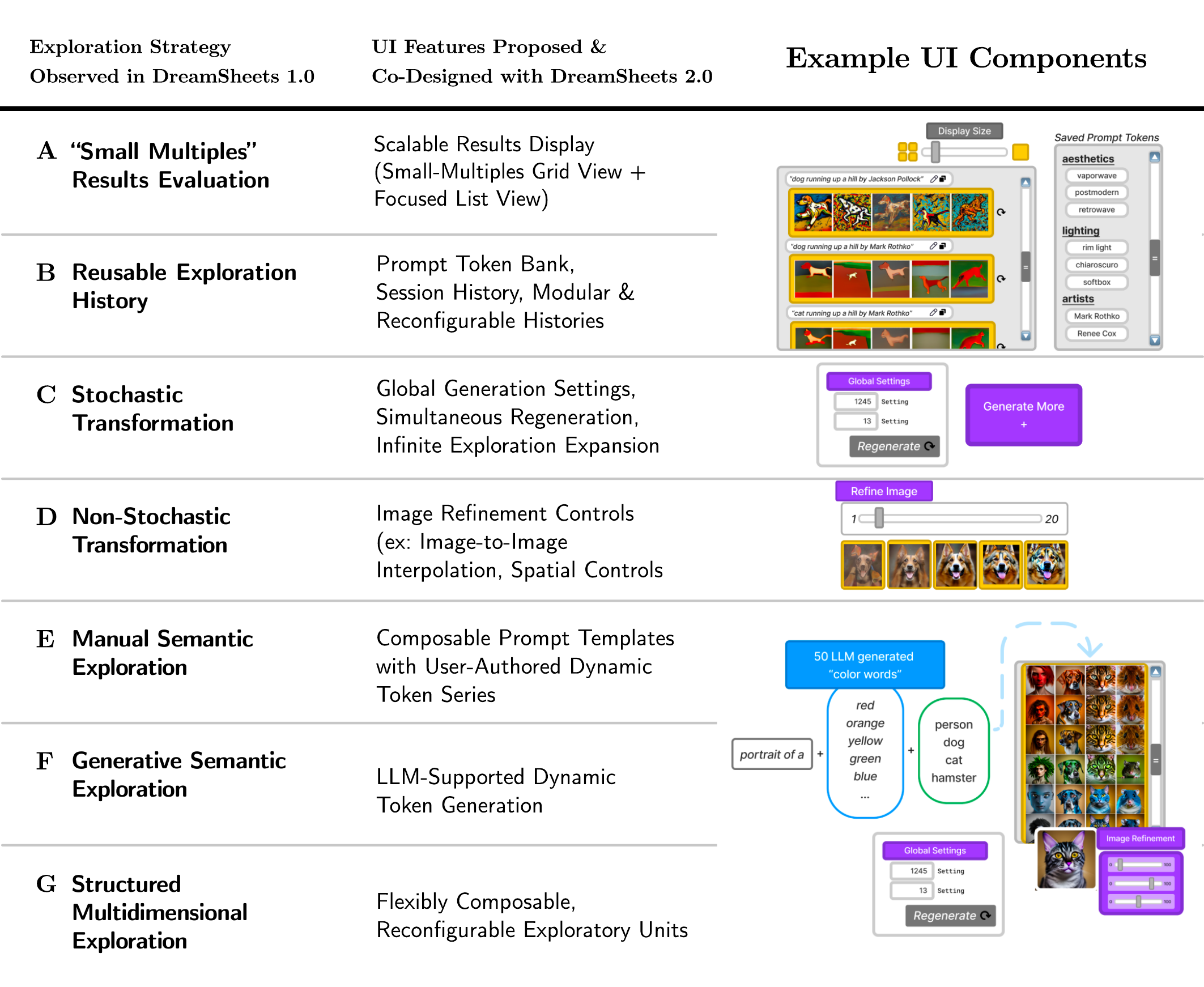}
    \caption{ By observing the exploratory sheet-systems developed by experts in \tool 1.0, we identified supportive strategies and structures. Through the \tool 2.0 Mock-Up Co-Design process, we elicited expert input and concrete suggestions to connect these ideas to generalizable UI concepts, which can be presented as composable ``units'' to support user-structured exploration in future interface designs. }
    \label{fig:uicomponents}
    \Description{A table mapping “Exploration Strategy Observed in DreamSheets 1.0” (column 1) to “UI Features Suggested \& Codesigned with DreamSheets 2.0” (column 2) with illustrative Example UI components in column 3. Each row is as follows: ROW 1: A. Small Multiples Results Evaluation, then: Scalable Results Display (Small-Multiples Grid View + Focused List View), a merged cell that spans rows 1 and 2 featuring the a smallmultiples image output display with a “Display Size” scaling slider and a prompt token bank. ROW 2: B. Reusable Exploration history, then: Prompt token bank, session history, reusable exploration histories, then: the 3rd cell as described in Row1. ROW3: C. Stochastic Transformation, then: Generation Settings, Simultaneous Regeneration, Infinite Exploration Expansion, then: UI components showing generation-wide settings and a “generate more” button, ROW4: D. Non-stochastic transformation, then: Image Refinement Controls(ex: Image-to-Image Interpolation, Spatial Controls), then: an image refinement slider that depicts a linear progressive effect on the series of images below the slider. ROW 5: E. Manual Semantic Exploration, then: Composable Prompt Templates with Customizable Dynamic Tokens, then a merged cell across the next 3 rows portraying a prompt template with “portrait of a” then a dynamic token indicating a series of 50 LLM generated color words, then a manually selected series of 5 subjects, combined for a large-scale series of results, which also indicate the sliders and generation wide controls for image refinement and parametric manipulation. ROW6: F. Generative Semantic Exploration, then: LLM-Suported Dynamic Token Generation, then the merged cell previously described. ROW7: G. Structured, Multidimensional Exploration, then: Flexibly Composable Reconfigurable Exploratory Units, then the merged cell previously described. 
}
\end{figure*}
To that end,} we designed the \tool 2.0 mockup to include two visual layout settings that users can freely toggle between: a ``small-multiples'' grid view (\ref{fig:mockuplarge}, left) and a focused list view(Fig. ~\ref{fig:mockuplarge}, right.)
``Scalable'' output display supports TTI users as they move between broad and focused evaluations of results. 

Participants used sheet and cell duplication to iteratively repurpose exploration history in \tool 1.0; this informed our decision to suggest features for revisiting prompt-history and exploration ``sessions'' in \tool 2.0. 


We proposed a prompt ``token bank'' system (Figure~\ref{fig:mockuplarge}, featured in Figure~\ref{fig:mockupsmall}, right) that would allow users to convert highlighted prompt text into a \textit{Saved Token} for reuse in future prompts. 
Tokens can be converted into \textit{dynamic} tokens for semantic explorations.

\edits{
\subsection{Supporting Exploration Breadth and Depth}
Participants utilized the structured ``infinite canvas’’ qualities of digital spreadsheets to iteratively expand explorations. 
Providing the user direct control over the ``scale'' of their generations allows them to flexibly expand and evaluate explorations. To this end, \tool 2.0 would continuously load more results on-demand, for potentially infinite scrolling. 
Cost may limit the feasibility of large-scale generations; offering cheaper ``low-fidelity'' generations and options to increase quality on-demand may be a potential solution.}
\edits{
We replicated the ``Power Cell'' strategy crafted by participants by providing options that would allow \tool 2.0 users to regenerate their exploration session by updating global hyperparameter settings. }

\edits{
We used the ``Explore this image...'' option to elicit participant ideas for additional ``image refinement controls'' beyond the \textit{classifier-free-guidance} manipulation available in \tool 1.0. 
}
\edits{
Suggestions included adding spatial conditioning controls (e.g. manipulating Poses or Edges; participants mentioned ControlNet ~\cite{zhang2023adding}), support for generative inpainting (as enabled by Muse~\cite{chang2023muse}), or image-to-image interpolation (as in SpaceSheets ~\cite{loh2018spacesheets}). Participants also suggested adding Image-to-Text transformations similar to the CLIP Interrogator~\cite{clip-interrogator2023} or Midjourney's \texttt{/describe} function~\cite{midjourney2023}, which could ``close the loop'' by allowing users to translate ideas back and forth between image and prompt space. 
}

\subsection{Supporting Semantic Exploration with Prompt Templates and Dynamic Tokens}
\edits{\tool 1.0 participants used cell concatenation to craft prompt templates with ``slots'' to select axes for structured semantic exploration, motivating the design of more supportive prompt-template features in our 2.0 Mock-Up.}  

\textit{Dynamic} tokens take the role of the ``slots'' used by participants to specify where to introduce text variation. 
Prompt variations are automatically generated with LLM-assistance, systematically combined, and populated across several columns of generated images. The LLM supporting this interaction is surfaced, providing users control over the text-generation, or the option to refuse its support altogether. Users can manually append, edit, and remove prompt words for each column of exploration, at will.  This component design meets \textit{Chang et al.}'s call for future interfaces to support prompt templates as standalone, interactive computational artifacts~\cite{chang_prompt_2023}. 
\edits{\subsection{Flexible Structures for User-Defined Multidimensional TTI Explorations}}
\tool 1.0 offered the flexibility of an infinite ``blank canvas'' with the trade-off of minimal structural support ``out of the box.'' Our expert participants were required to iteratively prototype systems custom to their creative ``styles'' and workflows.
Rather than designing for a particular workflow, \tool 2.0 suggests supportive structures while maintaining flexibility by presenting composable exploration features.
Users can effectively pursue simple prompt-input-image-output tests, or they can construct increasingly sophisticated multidimensional explorations. With configurable, reusable, and refusable components, users can compose targeted, iterative explorations of prompt-image space.  

\subsection{Participant Feedback}

 \edits{Participants reflected positively on the features presented in the \tool 2.0 Mock-Up, validating our adaptive interpretations of the structures they prototyped in \tool 1.0. The Mock-Up was developed and presented during the study; their feedback directly contributed to improving the design and identifying its most promising components. }
 The ability to save and recover exploration history for reuse in future explorations was a highlight: E4 considered the ``prompt token bank'' a direct upgrade to their current history keeping practices (saving prompts and useful stylistic modifiers into a text document). 
\begin{quote}
    ``If I can drag and drop a presaved dynamic chunk... I can fully focus on sculpting the prompt and being creative.'' (E4)
\end{quote}

The ``Save session'' feature prompted E1 and E5 to describe similar expected use cases – to save current state of the system, pausing their workflow to return at a later date, ideally supporting history-reuse. E5 expressed their preference to segment their process into a  ``generation'' stage (crafting prompts to generate thousands of images) and a ``evaluation'' stage (curating the selection of images), and described being able to pause and temporally separate these activities as a potential ``game-changer.'' 
\section{Limitations}
\edits{We acknowledge several limitations of our study. First, the particular hyperparameters exposed may differ between different TTI models; future work should investigate if our approach scales to higher-dimensional hyperparameter spaces. Second, more powerful techniques to guide image generation are emerging in research, such as ControlNet~\cite{zhang2023adding}, or Readout-Guidance~\cite{luo2023readout}, which require different forms of input to the TTI model. As participants requested, future work should investigate how \tool could be extended to support these approaches.}

\section{Conclusion}
Text-to-Image models challenge users to navigate vast, opaque design spaces on both sides---prompt input and image output. \tool provides a flexible, spreadsheet-based interface for users to author strategies to achieve creative goals, and facilitating sensemaking---developing through experience the language and working understanding needed to reliably steer image generations towards interesting outputs. Through two user studies, including an extended expert study, we observed challenges, tensions, and opportunities in the TTI prompt-exploration process. We utilized these insights to develop a UI mockup, improved with participant feedback, and suggesting features for future supportive TTI exploration interfaces. Finally, we considered the implications of supporting users' sensemaking in prompt-image space, and beyond. 
\section{Disclosure}
The authors used ChatGPT for minor copy editing tasks.
\section{Expert Artist Credits}
In consenting to participate in the study, participants gave explicit and informed consent for the data collected during the study, including their creative contributions, to be anonymously published in research findings. That said, the authors believe that the AI and research community should strive to credit artists when their work is used, according to their wishes. After acceptance, the authors reached out to the expert generative artist participants to collect their preferences with regard to remaining anonymous or receiving credit in the final publication.
\begin{enumerate}
    \item Expert Participant 1 (E1) is Stephen Young, a mixed media generative artist with 2 years of AI experience. He's worked with multiple GAN and diffusion models to create art available on his website \hyperlink{https://www.kyrick.art}{https://www.kyrick.art}, $@$kyrick.art on Threads and $@$kyrickyoung on X.com.
    \item Expert Participant 2 (E2) is Jeremy Torman, an interdisciplinary artist/musician that has been painting for over 20 years and using generative ai tools since 2016. He has used GANs, deepdream, style-transfer, vqgan+clip, jax diffusion, deforum, et al. He shares his work online $@$tormanjeremy on X/Twitter and $@$jeremy\_torman on Instagram.
    \item Expert Participant 3 (E3) is Seth Niimi, a multi-passionate creator who explores unique ways to combine tools, processes and ideas in pursuit of fascinating experiences. Seth can be found on TikTok and Instagram as $@$synaestheory.
    \item Expert Participant 4 (E4) chose to remain anonymous. 
    \item Expert Participant 5 (E5) is a Canadian generative artist known as sureai.i, with 2 years of experience in the TTI space. She has previously used a variety of tools, including Stable Diffusion and Midjourney, along with manual digital editing. Her work can be seen online at $@$sureailabs on X.com and $@$surea.i on Instagram.
\end{enumerate}
\begin{acks}
We would like to thank our reviewers for their time and effort; their suggestions were immensely helpful in improving the quality and clarity of this work. We would also like to thank our participants for their contributions to the studies. 
\end{acks}

\bibliographystyle{ACM-Reference-Format}
\bibliography{ref/dreamsheets,ref/ref,ref/codesign}
\end{document}